\documentclass[longauth]{aa}
\usepackage{graphicx}
\usepackage{txfonts}

\usepackage{multirow}

\usepackage[breaklinks=true,colorlinks=true,citecolor=blue,linkcolor=blue,urlcolor=blue]{hyperref}
\usepackage{tablefootnote}
\usepackage{natbib}
\usepackage{amsmath} 
\usepackage{amssymb}
\usepackage{mathtools}
\usepackage{microtype}
\usepackage[T1]{fontenc}
\usepackage[utf8]{inputenc}
\usepackage[dvipsnames]{xcolor}
\usepackage{array}
\usepackage{makecell}
\usepackage{multirow}
\usepackage{siunitx}
\usepackage{tabularx}
\usepackage{caption}

\usepackage[caption=false]{subfig}
\usepackage{verbatim}
\usepackage{xspace}

\usepackage{hyperref}
\newcommand{\grav}{${\rm GRAVITY}$\xspace} 
\def\kms{\rm\, km\,s^{-1}}
\def\hagn{\mbox{{\sc \small Horizon-AGN}}}
\def\msun{\,{\rm M_\odot}}
\def\kms{\,{\rm km\,s^{-1}}}

\begin{document} 
\title{Searching for Nearby Supermassive Black Hole Binaries with GRAVITY}
\authorrunning{GRAVITY+ Collaboration}
\author{GRAVITY Collaboration: K.~Abd El Dayem\inst{\ref{ref:LESIA}}
\and N.~Aimar\inst{\ref{ref:Porto},\ref{ref:CENTRA}}
\and A.~Berdeu\inst{\ref{ref:ESOGarching},\ref{ref:LESIA}}
\and J.-P.~Berger\inst{\ref{ref:IPAG}}
\and G.~Bourdarot\inst{\ref{ref:MPE}}
\and W.~Brandner\inst{\ref{ref:MPIA}}
\and Y.~Cao\inst{\ref{ref:MPE}}
\and C.~Correia\inst{\ref{ref:Porto},\ref{ref:CENTRA}}
\and Q.~D'Amato\inst{\ref{ref:inafroma}}
\and R.~Davies\inst{{\ref{ref:MPE}}}
\and D.~Defr{\`e}re\inst{\ref{ref:KUL}}
\and A.~Drescher\inst{\ref{ref:IPAG}}
\and F.~Eisenhauer\inst{\ref{ref:MPE},\ref{ref:TUM}}
\and M.~Fabricius\inst{\ref{ref:MPE}}
\and H.~Feuchtgruber\inst{\ref{ref:MPE}}
\and S.~Flesch\inst{\ref{ref:MPE}}
\and N.~M.~F{\"o}rster~Schreiber\inst{\ref{ref:MPE}}
\and A.~Foschi\inst{\ref{ref:LESIA}}
\and Q.~Fournier\inst{\ref{ref:MPE}}
\and P.~Garcia\inst{\ref{ref:Porto},\ref{ref:CENTRA}}
\and R.~Garcia~Lopez\inst{\ref{ref:UCD}}
\and R.~Genzel\inst{\ref{ref:MPE},\ref{ref:Berkeley}}
\and S.~Gillessen\inst{\ref{ref:MPE}}
\and F.~Gont{\'e}\inst{\ref{ref:ESOGarching}}
\and X.~Haubois\inst{\ref{ref:ESOChile}}
\and S.F.~H{\"o}nig\inst{\ref{ref:Southampton}}
\and L.~C.~Ho\inst{\ref{ref:kavlichina},\ref{ref:unichina}}
\and M.~Houll{\'e}\inst{\ref{ref:IPAG}}
\and S.~Joharle\inst{\ref{ref:MPE}}
\and J.~Kammerer\inst{\ref{ref:ESOGarching}}
\and A.~Kaufer\inst{\ref{ref:ESOChile}}
\and P.~Kervella\inst{\ref{ref:LESIA},\ref{ref:FCLA}}
\and J.~Kolb\inst{\ref{ref:ESOGarching}}
\and L.~Kreidberg\inst{\ref{ref:MPIA}}
\and L.~Labadie\inst{\ref{ref:Cologne}}
\and S.~Lacour\inst{\ref{ref:LESIA},\ref{ref:ESOGarching}}
\and O.~Lai\inst{\ref{ref:Lagrange}}
\and R.~Laugier\inst{\ref{ref:KUL}}
\and J.-B.~Le~Bouquin\inst{\ref{ref:IPAG}}
\and J.~Leftley\inst{\ref{ref:ESOChile}}
\and R.~Li\inst{\ref{ref:MPE}}
\and K.~Li\inst{\ref{ref:Flatiron}}
\and B.~Lopez\inst{\ref{ref:Lagrange}}
\and D.~Lutz\inst{\ref{ref:MPE}}
\and F.~Mannucci\inst{\ref{ref:inafarcetri}}
\and F.~Mang\inst{\ref{ref:MPE}}
\and A.~M{\'e}rand\inst{\ref{ref:ESOGarching}}
\and F.~Millour\inst{\ref{ref:Lagrange}}
\and M.~Montarg{\`e}s\inst{\ref{ref:LESIA}}
\and N.~Moruj{\~a}o\inst{\ref{ref:Porto},\ref{ref:CENTRA}}
\and H.~Nowacki\inst{\ref{ref:Lagrange}}
\and M.~Nowak\inst{\ref{ref:LESIA}}
\and S.~Oberti\inst{\ref{ref:ESOGarching}}
\and J.~Osorno\inst{\ref{ref:LESIA}}
\and T.~Ott\inst{\ref{ref:MPE}}
\and C.~Paladini\inst{\ref{ref:ESOChile}}
\and S.~Pappert\inst{\ref{ref:MPE}}
\and T.~Paumard\inst{\ref{ref:LESIA}}
\and K.~Perraut\inst{\ref{ref:IPAG}}
\and G.~Perrin\inst{\ref{ref:LESIA}}
\and R.~Petrov\inst{\ref{ref:Lagrange}}
\and P.O.~Petrucci\inst{\ref{ref:IPAG}}
\and N.~Pourr{\'e}\inst{\ref{ref:IPAG}}
\and S.~Rabien\inst{\ref{ref:MPE}}
\and D.C.~Ribeiro\inst{\ref{ref:MPE}}
\and S.~Robbe-Dubois\inst{\ref{ref:Lagrange}}
\and M.~Sadun~Bordoni\inst{\ref{ref:MPE}}
\and J.~Sanchez-Bermudez\inst{\ref{ref:UNAM}}
\and J.~Sauter\inst{\ref{ref:MPIA}}
\and M.~Scialpi\inst{\ref{ref:MPE},\ref{ref:UNITN},\ref{ref:UNIFI},\ref{ref:inafarcetri}}
\thanks{Corresponding author: martina.scialpi@inaf.it}
\and J.~Scigliuto\inst{\ref{ref:Lagrange}}
\and J.~Shangguan\inst{\ref{ref:kavlichina},\ref{ref:MPE}}
\and T.~Shimizu\inst{\ref{ref:MPE}}
\and F.~Soulez\inst{\ref{ref:CRAL}}
\and C.~Straubmeier\inst{\ref{ref:Cologne}}
\and E.~Sturm\inst{\ref{ref:MPE}}
\and M.~Subroweit\inst{\ref{ref:Cologne}}
\and C.~Sykes\inst{\ref{ref:Southampton}}
\and L.J.~Tacconi\inst{\ref{ref:MPE}}
\and P.~Th{\'e}venet\inst{\ref{ref:LESIA}}
\and I.~Urso\inst{\ref{ref:LESIA}}
\and F.~Vincent\inst{\ref{ref:LESIA}}
\and J.~Woillez\inst{\ref{ref:ESOGarching}}
\and G.~Zins\inst{\ref{ref:ESOChile}}
}

\institute{
LESIA, Observatoire de Paris, Université PSL, CNRS, Sorbonne Université, Université Paris Cité, 5 Place Jules Janssen, 92195 Meudon, France
\label{ref:LESIA}
\and
Faculdade de Engenharia, Universidade do Porto, Rua Dr. Roberto Frias, 4200-465 Porto, Portugal
\label{ref:Porto}
\and
CENTRA - Centro de Astrofísica e Gravitação, IST, Universidade de Lisboa, 1049-001 Lisboa, Portugal
\label{ref:CENTRA}
\and
European Southern Observatory, Karl-Schwarzschild-Straße 2, 85748 Garching, Germany
\label{ref:ESOGarching}
\and
Univ. Grenoble Alpes, CNRS, IPAG, 38000 Grenoble, France
\label{ref:IPAG}
\and
Max-Planck-Institut für extraterrestrische Physik (MPE), Gie{\ss}enbachstra{\ss}e 1, 85748 Garching, Germany
\label{ref:MPE}
\and
Max-Planck-Institut für Astronomie (MPIA), Königstuhl 17, 69117 Heidelberg, Germany
\label{ref:MPIA}
\and
INAF - Istituto di Astrofisica e Planetologia Spaziali, Via del Fosso del Cavaliere, 00133 Roma, Italy
\label{ref:inafroma}
\and
Institute of Astronomy, KU Leuven, Celestijnenlaan 200D, 3001 Leuven, Belgium
\label{ref:KUL}
\and
Department of Physics, Technical University of Munich, 85748 Garching, Germany
\label{ref:TUM}
\and
School of Physics, University College Dublin, Belfield, Dublin 4, Ireland
\label{ref:UCD}
\and
Departments of Physics and Astronomy, University of California, Berkeley, Berkeley, CA, USA
\label{ref:Berkeley}
\and
European Southern Observatory, Alonso de Córdova 3107, Casilla 19001, Vitacura, Santiago, Chile
\label{ref:ESOChile}
\and
Department of Physics and Astronomy, University of Southampton, Southampton, UK
\label{ref:Southampton}
\and
The Kavli Institute for Astronomy and Astrophysics, Peking University, Beijing 100871, China
\label{ref:kavlichina}
\and
Department of Astronomy, School of Physics, Peking University, Beijing 100871, People's Republic of China
\label{ref:unichina}
\and
I. Physikalisches Institut, Universität zu Köln, Zülpicher Straße 77, 50937 Cologne, Germany
\label{ref:Cologne}
\and
Université Côte d'Azur, Observatoire de la Côte d'Azur, CNRS, Laboratoire Lagrange, Nice, France
\label{ref:Lagrange}
\and
Center for Computational Astrophysics, Flatiron Institute, 162 Fifth Avenue, New York, NY 10010, USA
\label{ref:Flatiron}
\and
INAF - Osservatorio Astrofisico di Arcetri, Largo E. Fermi 5, 50125 Firenze, Italy
\label{ref:inafarcetri}
\and
Universidad Nacional Autonoma de Mexico, Instituto de Astronom\'ia, A. P. 702-64, 04510, Ciudad de Mexico, Mexico 
\label{ref:UNAM}
\and
University of Trento, Via Sommarive 14, I-38123 Trento, Italy
\label{ref:UNITN}
\and
Università di Firenze, Dipartimento di Fisica e Astronomia,
via G. Sansone 1, 50019 Sesto Fiorentino (FI), Italy
\label{ref:UNIFI}
\and
Univ. Lyon, Univ. Lyon 1, ENS de Lyon, CNRS, Centre de Recherche Astrophysique de Lyon UMR5574, F-69230, Saint Genis-Laval, France
\label{ref:CRAL}
\and
French-Chilean Laboratory for Astronomy, IRL 3386, CNRS and Universidad de Chile, Casilla 36-D, Santiago, Chile
\label{ref:FCLA}
}

\date{}
\abstract{We present a systematic interferometric search for close supermassive black hole (SMBH) binaries in 20 nearby type 1 active galactic nuclei (AGN) observed with VLTI/\grav.
We investigate parsec-scale asymmetric emission by modeling the squared visibilities and closure phases with a single circular Gaussian component plus a background point source for the central AGN, and searching for additional compact companions through grid-based binary models. The robustness of the search is validated using injection--recovery simulations and point-spread-function diagnostics.

Several features are identified in the initial significance maps, but only one candidate out of 20 shows a persistent compact asymmetry, corresponding to an observed incidence of $5\%$. If NGC~3783 is interpreted as a binary AGN, we derive a Poissonian one-sided $1\sigma$ upper limit of $f_{\rm bin}<16.5\%$ for \grav-detectable binaries among bright nearby AGN ($K<12.5$) at separations of $\sim1$--$100$~pc. Conversely, if the observed asymmetry in NGC~3783 is associated with asymmetric circumnuclear dust emission, the corresponding upper limit on the binary fraction is $f_{\rm bin}<9.2\%$.

We compare these results with predictions from the Horizon-AGN simulation, finding comparable binary fractions. We also noted that the expected fraction strongly depends on the assumed distribution of individual SMBH accretion rates within binaries, which is currently unresolved in cosmological models. 

Our results demonstrate the potential of infrared interferometry to identify parsec-scale binary AGN candidates and motivate future \mbox{\grav}+ observations to expand the sample and follow up known candidates, helping to confirm their nature and distinguish genuine binaries from asymmetric nuclear structures.
}

   \maketitle

\section{Introduction}
In the framework of the $\Lambda$CDM cosmological model, structures in the Universe form hierarchically, with galaxies growing through mergers and accretion \citep{White78, Begelman80}. Most massive galaxies host a central supermassive black hole (SMBH), which can grow both by accreting gas and through coalescence with other SMBHs during galaxy mergers \citep{DiMatteo05, Kormendy_Ho_13}.

Galaxy mergers lead to a sequence of evolutionary phases for supermassive black holes (SMBHs), which may be observed as active galactic nuclei (AGN) when they are actively accreting.
The first stage is represented by dual AGN, in which the two SMBHs are hosted by two distinct merging galaxies and are typically observed at projected separations from several tens of kiloparsecs down to few kpc. As the merger progresses and the galaxies coalesce into a single remnant, the system enters the so-called Cosmic Duet phase, characterized by two active nuclei at separations below $\sim$7 kpc within the same galaxy. 
As dynamical interactions with stars, gas, and dark matter reduce their separation driving the hardening of the binary system at parsec scales which eventually form a gravitationally bound system at sub-parsec separations \citep{derosa20,Krause25}. 
These compact binaries are expected to merge, representing some of the most promising multi-messenger sources \citep{Kelley21, charisi22}, as they can emit both bright electromagnetic (EM) signatures \citep{Bogdanovic21,DOrazioCharisi23} and low-frequency gravitational waves (GWs), detectable at nanohertz frequencies by pulsar timing arrays (PTAs) \citep{Burke-Spolaor19, EPTACollaboration2023} or at millihertz frequencies by future space-based observatories such as LISA \citep{Agazie2023a, AmaroSeoane2023, Colpi2024arxiv}.

Observations of dual AGN and Cosmic Duets provide key constraints on the frequency of galaxy mergers, the properties of their host galaxies, and the accretion behavior of SMBHs during early merger stages \citep{koss12, Rosas-Guevara2019, derosa20, Volonteri22}. A significant fraction of known systems has been discovered serendipitously \citep[e.g.,][]{Junkkarinen01, Glikman23, Maiolino2024dual, Ubler2024dual, Matsuoka2024, Perna2025dual,Zamora25}, reflecting both their intrinsic rarity and the strong angular-resolution requirements needed to resolve dual nuclear activity. 
More recently, systematic searches leveraging wide-field surveys such as DESI, Gaia, and Euclid \citep{DESI25_arxive, Gaia_EDR3_1, Euclid_Overview} have begun to significantly increase the number of candidates \citep{Mannucci22,Chen22a,desi_dual,Ulivi2025arXiv}.
At these separations (from several kiloparsecs down to a few hundred parsecs), dual AGN can be confirmed through optical and near-infrared spectroscopy \citep{Mannucci22,Mannucci23,Scialpi24,scialpi25_arxive,Lemon17,Shen23b,Lemon23,mezcua26}, although selection effects due to obscuration and contamination from star formation remain significant limitations . Multi-wavelength follow-up, in particular X-ray and radio observations, plays a crucial role in confirming the presence of two accreting SMBHs mitigating biases inherent to optical selection and studies different properties \citep{Fu15_a,Fu15_b,DeRosa2023,Battistini26}.

At progressively smaller separations, the identification of binary AGN becomes increasingly challenging. Early-stage candidates have been identified using double-peaked emission lines and imaging of merging nuclei, although these methods are often degenerate with kinematic or disc-related effects in single AGN \citep{Wang09_binary_doublepeak, Liu10_binary_doublepeak, Severgnini18_binary_doublepeak, Koss2018}. At sub-parsec separations, spectroscopic radial-velocity monitoring has been proposed to trace the orbital motion of the black holes through systematic shifts of broad emission lines \citep{Runnoe17, Runnoe25}. An alternative approach is provided by time-domain photometry, where periodic or quasi-periodic variability in AGN light curves has been proposed as a potential signature of binary SMBHs \citep{Charisi16, Charisi18, charisi22}. However, periodicity searches are strongly affected by the intrinsic stochastic variability of single AGN, which can produce apparent periodic signals, particularly over observational baselines spanning only a few cycles. Moreover, the expected light-curve morphology depends on the physical mechanism responsible for the modulation, making a simple periodicity criterion insufficient to establish a binary interpretation. These limitations have motivated the use of increasingly stringent statistical tests and complementary electromagnetic diagnostics. To date, no statistically significant sample of confirmed sub-parsec SMBH binaries exists.

Interferometric techniques provide another promising approach. Differential interferometry can probe wavelength-dependent photocentre shifts across broad emission lines, tracing the kinematics of unresolved SMBH binaries. \cite{Kovacevic20} showed that differential-phase profiles can encode the orbital configuration of close binaries, while also highlighting degeneracies with single SMBHs hosting clouds on elliptical orbits. Very Long Baseline Interferometry (VLBI) has also enabled the detection of a few exceptional systems at parsec scales, such as the radio galaxy 0402+379, with a projected separation of $\sim$7.3 pc \citep{Rodriguez06}. VLBI primarily probes compact non-thermal radio emission and therefore selects a population complementary to that targeted by near-infrared interferometry. In contrast, VLTI/\grav\ probes the thermal near-infrared emission from hot dust around the AGN, enabling the search for close SMBH pairs in a different AGN population.

VLTI/\grav is a near-infrared K-band interferometer that achieves enhanced sensitivity and $(u,v)$-plane coverage by coherently combining the light from the four Unit Telescopes, operating as a six-baseline array \citep{Gravity+17_1}. Its continuum observations reach an angular resolution of $\sim$3--5 mas, corresponding to physical scales of $\sim$1--100 pc. Unlike VLBI, however, \grav probes the thermal emission from the innermost regions of active galaxies, tracing the hot dust surrounding the nucleus in low-redshift AGN and the accretion disk emission in higher-redshift systems. Furthermore, \grav has demonstrated remarkable observational efficiency: only about one hour of integration is sufficient to resolve and measure the size of the hot-dust continuum in a nearby AGN \citep{gravity_taro_NGC3783}.
Previous \grav studies exploited the interferometric visibility amplitudes to constrain the characteristic size of the continuum-emitting region \citep{Cao24,gravity20,dexter_gravity20}. In this work, instead, we focus on the closure phases, which are directly sensitive to departures from point symmetry and therefore provide a powerful probe of binary structures in the hot-dust emission.

In this work, we present a new analysis on VLTI/\grav data of a sample of 20 type 1 AGN published in \cite{Cao24,gravity20} and \cite{dexter_gravity20} at redshift below 0.25. By systematically analyzing deviations from point symmetry, we search for candidate binary signals and place stringent upper limits on secondary components. Our study demonstrates the power of high-precision interferometry to probe the earliest stages of SMBHs pairing and highlights how the upcoming \grav+ upgrade will extend these capabilities to fainter targets and smaller flux ratios, enabling systematic searches across a wide range of AGN luminosities and redshifts.
The paper is organized as follows. In Sect.~\ref{sec:cp}, we introduce the closure phase and its application to the identification of binary AGN candidates with \grav. Section~\ref{sec:dataset} presents the observations while Section~\ref{sec:analysis} describe the interferometric analysis to fit the central source and find any additional emission through a companion search. The results and the inferred binary AGN fraction are presented in Sects.~\ref{sec:binarycandidates} and \ref{sec:binary_fraction}, respectively. The Horizon-AGN model is introduced in Sect.~\ref{HAGN}, and the comparison between the observational results and theoretical predictions is discussed in Sect.~\ref{sec:comparison}.
We adopted a flat $\Lambda$CDM cosmology with parameters from \cite{Planck2020}, with $H_{\rm 0}=67.8\ {\rm km\ s^{-1}\ Mpc^{-1}}$ and $\Omega_{\rm m,0}=0.38$. 

\section{The role of closure phase}
\label{sec:cp}
The closure phase (T3PHI) is defined as the sum of the interferometric phases measured on the three baselines forming a closed triangle, $T3PHI = \phi_{12} + \phi_{23} + \phi_{31}$. It is a robust observable in optical/infrared interferometry, as it is immune to atmospheric phase fluctuations and instrumental phase errors, making it particularly sensitive to deviations from point-symmetric brightness distributions \citep{Eisenhauer23}. In the context of local AGN observations with VLTI/\grav, closure phase measurements therefore provide a direct diagnostic of asymmetries in the hot dust continuum emission that dominates the near-infrared flux \citep{Monnier03}.

For a centro-symmetric brightness distribution, the closure phase is expected to be close to zero, with deviations limited by the measurement uncertainties. A significant non-zero closure phase indicates a departure from centro-symmetry and can arise from a variety of physical configurations, including clumpy or non-axisymmetric dust emission, asymmetric illumination, or the presence of an additional compact source such as a secondary accreting SMBH.

\begin{figure}
    \centering
    \includegraphics[width=0.9\linewidth]{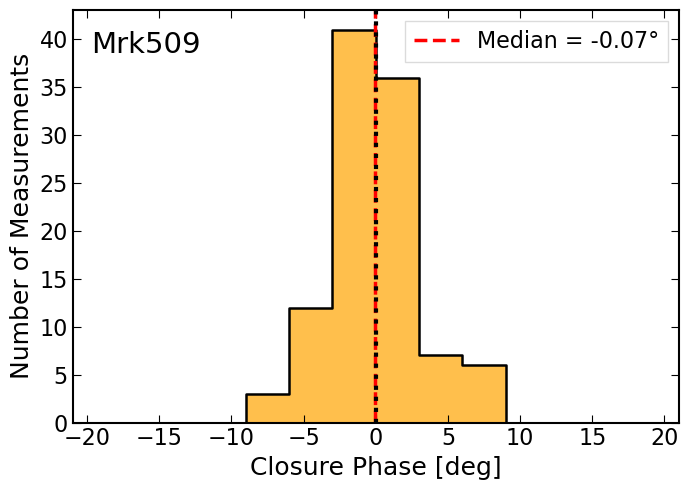}
    \includegraphics[width=0.9\linewidth]{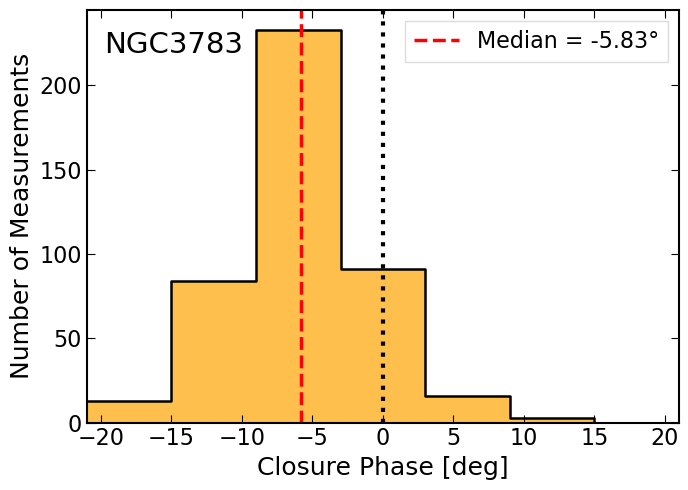}
    \caption{Multiple measurements of the closure phase for Mrk 509 (upper panel) and NGC 3783 (lower panel), with median values of $-0.07^\circ$ and $-5.83^\circ$, respectively (red lines). The zero phase is indicated by the black line.
 }
    \label{fig:cp}
\end{figure}

Simulations and previous \grav observations (e.g. NGC 3783; \cite{gravity_taro_NGC3783}) show that even a weak secondary source or asymmetric hot dust emission can produce measurable closure-phase offsets of several degrees.
Figure~\ref{fig:cp} illustrates the closure-phase distributions measured for two representative sources in our sample, Mrk 509 and NGC 3783. Mrk 509 shows closure phases distributed around $0^\circ$, with a median of $-0.07^\circ$ and values statistically consistent with zero (one-sample Wilcoxon test: $p = 0.60$), whereas NGC 3783 exhibits a broader distribution with a median of $-5.83^\circ$ and a highly significant deviation from zero (one-sample Wilcoxon test: $p = 3.9\times10^{-51}$). These examples illustrate both the near-zero closure phases expected for a predominantly symmetric emission distribution as in Mrk 509 and the statistically measurable offsets that can arise from compact asymmetric emission like in NGC 3783.

The interferometric measurements probe milli-arcsec angular scales, corresponding to projected physical separations of approximately $\sim$1--$\sim$100~pc over the redshift range of our sample. These scales are substantially smaller than those accessible to conventional imaging or spectroscopic observations, allowing us to probe the compact nuclear region where closely separated accreting black holes may reside.

We exploit this sensitivity to asymmetries to perform a systematic search for compact binary AGN candidates in a large, homogeneous sample with a well-defined selection function. We model the interferometric observables as a function of projected separation and flux ratio to test whether deviations from a symmetric single-component model can be explained by an additional compact source.
Closure phases provide the primary constraint on departures from centro-symmetry, while squared visibilities ($V^2$) quantify how well the source is spatially resolved at a given spatial frequency, thereby constraining the characteristic size and spatial distribution of the emission. 
The two observables are therefore jointly used throughout the analysis to identify and characterize candidate compact companions.

\subsection{Sensitivity regime and binary signatures}

To quantify the parameter space accessible to \grav, we explore the expected closure phase response for binary configurations consisting of a dominant central source and a secondary compact component, with flux ratios between 0.1 and 1 and projected separations spanning 1--40 mas, corresponding to the angular separation range probed in our \grav observations.
The simulated closure-phase response for a range of separations and orientations is presented in Appendix~\ref{app:closure_phase_simulations}.
In this regime, the closure phase signal transitions from values consistent with zero, when the companion is either too faint or unresolved, to strongly angle-dependent signatures characteristic of asymmetric emission in the continuum.
In the unresolved regime, closure phases remain negligible within instrumental uncertainties, while partially resolved configurations produce non-zero, baseline-dependent signals that can reach amplitudes of several degrees, consistent with \grav sensitivity limits demonstrated in previous observations \citep{gravity_taro_NGC3783}.
In particular, for unequal-flux configurations, the closure phase amplitude generally increases with increasing flux ratio between the two components, while its characteristic oscillatory pattern becomes more pronounced for larger separations of the secondary. In the limiting case of an equal-flux binary, however, the brightness distribution becomes centro-symmetric and the closure phase vanishes. In this limiting case, the absence of a closure-phase signal does not imply the absence of a binary signature in the squared visibilities, which can still retain information on the separation of the two components.
As a result, the closure phase encodes the geometry of the system through its dependence on flux ratio, separation, and orientation, enabling partial reconstruction of the binary configuration from baseline-dependent variations in $(u,v)$ space. This methodology enables \grav to constrain the occurrence rate of binary AGN in the $\sim$1–100 pc regime.

\section{Sample definition, observations, and data reduction}
\label{sec:dataset}
We select a sample of low-redshift AGN observed with VLT/\grav in both single-field on-axis and off-axis modes. The sample includes the targets presented in \cite{dexter_gravity20} and \cite{Cao24}, thereby providing a homogeneous set of \grav observations obtained with comparable instrumental configurations but targeting different aspects of the nuclear emission. The targets were originally selected for their suitability for high-angular-resolution interferometric studies of the AGN broad-line region and/or hot-dust continuum, rather than specifically for searches for binary companions. This provides a well-defined parent sample in which to systematically assess the sensitivity of \grav to compact asymmetric emission.
The sample spans $0.002 \lesssim z \lesssim 0.25$ and covers a wide range of nuclear activity, with bolometric luminosities $\log L_{\mathrm{bol}} \sim 42.3$--$47.0$ erg s$^{-1}$, black hole masses $\log (M_{\mathrm{BH}}/M_\odot) \sim 5.7$--$8.8$, and Eddington ratios ranging from $\log \lambda_{\mathrm{Edd}} \sim -2$ to mildly super-Eddington values.

The observations were obtained using two single-field configurations.
In the on-axis mode, the incoming light is split approximately equally between the science channels and the fringe tracker (FT), enabling spectro-interferometric observations. This configuration was primarily adopted for targets requiring spectro-astrometric measurements of the broad-line region, typically at medium spectral resolution ($R=500$), and observations were often repeated over multiple epochs. The FT operates at $\sim300$ Hz using six spectral channels across the $K$ band, providing real-time phase referencing and enabling coherent integration of the science data. In the off-axis mode, instead, the full target flux is directed to the FT, maximizing the sensitivity to faint sources. This configuration was primarily used for continuum-focused observations, allowing the hot-dust emission to be characterized, typically within a single $\sim1$ hour observing epoch. Although the limiting magnitude remains set by the FT performance, the higher throughput of the off-axis configuration enables observations of fainter AGN than in the on-axis mode.

In this work, we use exclusively the FT continuum data products of observations obtained in both on-axis and off-axis configurations, without distinguishing between the two observing modes.

\subsection{Data Reduction}
The data were reduced using the standard \grav pipeline \citep{Lapeyrere14,Gravity+17_1}, with default settings for the extraction of FT visibilities and closure phases. All valid detector integrations were retained without additional selection based on fringe-tracker signal-to-noise ratio or V-factor. 
We adopt a group-delay-based selection to mitigate coherence losses associated with adaptive optics performance and fringe-tracking fluctuations. We require $|{\rm GDELAY}| < 3\mu\mathrm{m}$, following \cite{dexter_gravity20}, to reject low-quality frames. The same selection is applied consistently to both science and calibrator data.
After the uncalibrated reduction, the data are calibrated with an appropriate calibrator whose differential phase values are consistent with zero to remove residual instrumental effects.

The calibrated interferometric observables, including visibilities and closure phases, were averaged using inverse-variance weighting \citep{Gravity18} and stored in individual OIFITS files, which were then used for model fitting. Measurements with null projected baselines (u=v=0), arising from invalid telescope configurations, were discarded. For targets observed as part of the \grav BLR Large Programme (1103.B-0626, PI: Sturm), multiple observing blocks spanning one or more nights are available. These data were analyzed both on a per-night basis, to assess the robustness of candidate detections, and as a combined dataset to maximize the (u,v)-coverage and signal-to-noise ratio. Targets with only one or a few observing blocks (0109.B-0270, PI: Davies) were analyzed using all available data. No additional averaging between observing blocks was performed, as all calibrated OIFITS files were fitted simultaneously within {\tt PMOIRED}.

\section{Data analysis: binary candidate search}
\label{sec:analysis}

\begin{figure*}
    \centering
    \includegraphics[width=0.79\linewidth]{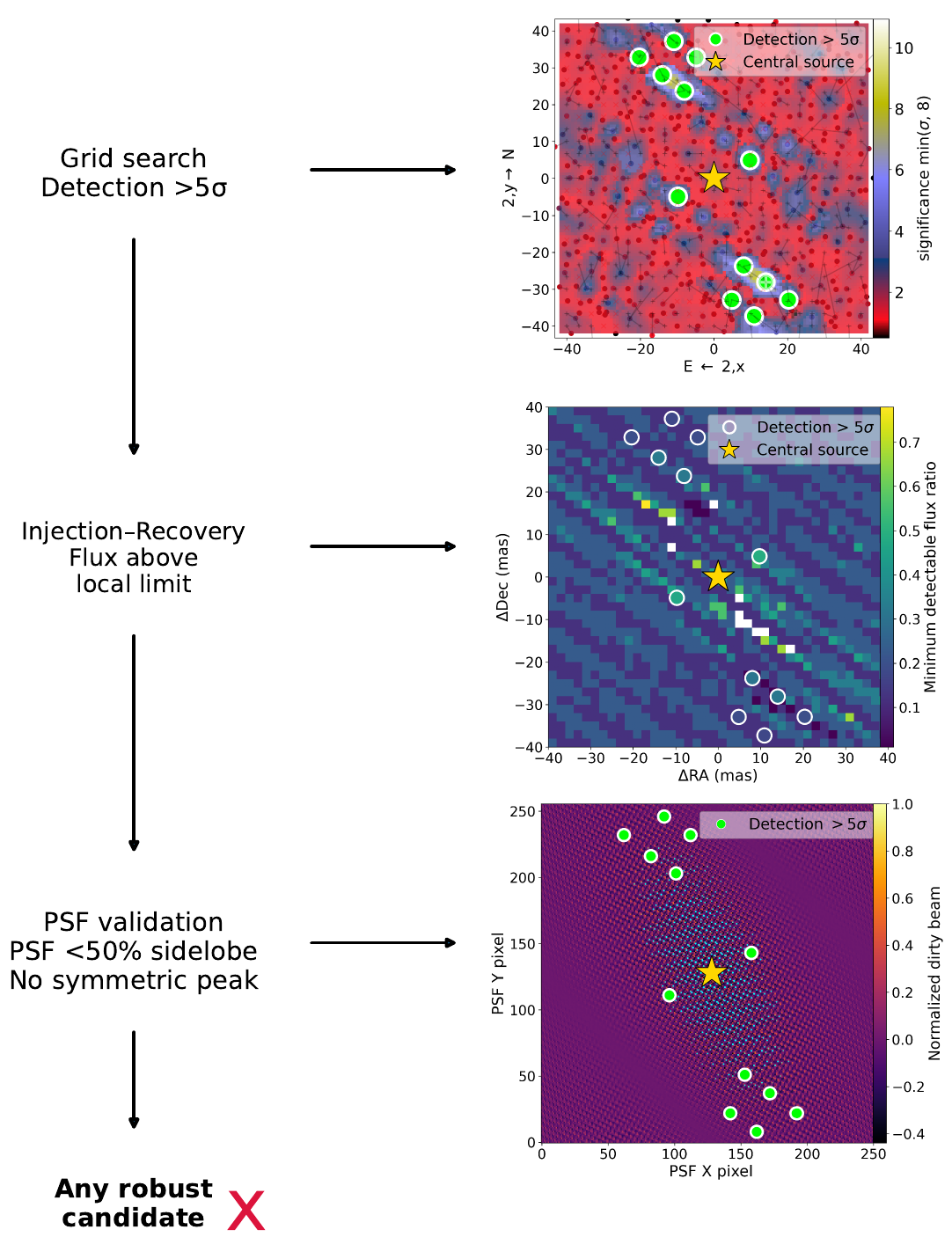}
    \vspace{8mm}
    \caption{Example of the candidate validation workflow for NGC 7603. The yellow star marks the position of the central source. The panels illustrate the successive steps of the candidate validation procedure. The first panel shows the significance map obtained from the grid search, with the colour bar indicating the detection significance in units of $\sigma$. Candidate positions exceeding the adopted $5\sigma$ threshold are highlighted by green circles. The second panel shows the injection--recovery detection-limit map, with the colur scale indicating the minimum detectable flux ratio at each position. Candidate positions are overlaid as open circles, with their colour encoding their measured flux ratio; candidates are retained only if their measured flux ratio exceeds the local detection limit. 
    The third panel shows the PSF-based diagnostic. The colour scale represents the absolute normalized PSF response, while the cyan contours delimit the regions where the PSF response exceeds 50\% of the maximum sidelobe amplitude outside the central PSF core. Candidate positions falling within these contoured regions, or showing a symmetric counterpart consistent with the PSF structure, are rejected as likely PSF artefacts. 
    In this example, no candidate survives all validation steps.}
    \label{fig:workflow}
\end{figure*}

We perform a systematic analysis of the VLTI/\grav\ interferometric data both on a per-night basis and using the combined datasets, in order to maximize spatial resolution and sensitivity to faint companions.
The analysis consists of two main steps followed by two validation and diagnostic procedures. Each step is performed using both $V^2$ and T3PHI to exploit the full interferometric information. 
First, we model each dataset with a central source consisting of a circular Gaussian component, representing the resolved hot-dust emission, and a co-located unresolved point source representing the central AGN. 
Second, we search for candidate secondary components in the residuals of the fit. 
We then assess the reliability of the detections through injection--recovery simulations and PSF artifact tests.
This procedure is implemented with the Python tool {\tt PMOIRED} \citep{PMOIRED22}, which enables parametric modeling of spectro-interferometric data using simple geometric components such as disks, rings, and Gaussians, and provides routines for grid-based fitting, bootstrapping, and detection-limit estimation. The three steps of the workflow are described in the following sections, with an example of the procedure applied to NGC~7603 shown in Fig.~\ref{fig:workflow}.

\subsection{Step 1: Central Gaussian fitting}
\label{sec:fwhm}
We first fit the source with a central component consisting of a circular Gaussian representing the resolved hot-dust emission, with an optimized full width at half maximum ($\mathrm{FWHM} = 0.62 \mathrm{mas}$), and a co-located unresolved point source representing the central AGN. The model also includes a resolved flux component, whose fractional contribution $f_{\rm res}$ is fitted simultaneously with the Gaussian FWHM and constrained to lie between 0 and 1.

We use a circular Gaussian as the reference model for the central resolved emission, following previous GRAVITY studies of nearby type-1 AGN \citep{gravity20}. We also tested elliptical Gaussian models for the central component, allowing the inclination and position angle to vary, and found that the additional degrees of freedom do not generally provide a significant improvement in the Bayesian Information Criterion (BIC), which penalizes models with a larger number of free parameters. Moreover, the inferred central sizes and unresolved flux fractions remain consistent within the uncertainties. We therefore adopt the circular Gaussian as the reference model throughout the analysis.
This model has two free parameters: the Gaussian Full Width at Half Maximum (FWHM) and the unresolved flux fraction (res,f), with the Gaussian flux given by the remaining fraction. We simultaneously fit the squared visibilities and closure phases to obtain a reference axisymmetric model that constrains the continuum size and unresolved emission before searching for binary companions.
Parameter uncertainties are estimated via bootstrap resampling of the interferometric observables, producing a distribution of best-fit solutions and thus robust confidence intervals on both the FWHM and flux fraction. 
The resulting solution provides a robust reference for the central continuum emission and serves as a baseline for the search for additional substructures or faint companions.
\begin{figure}[h!]
    \centering
    \includegraphics[width=1\linewidth]{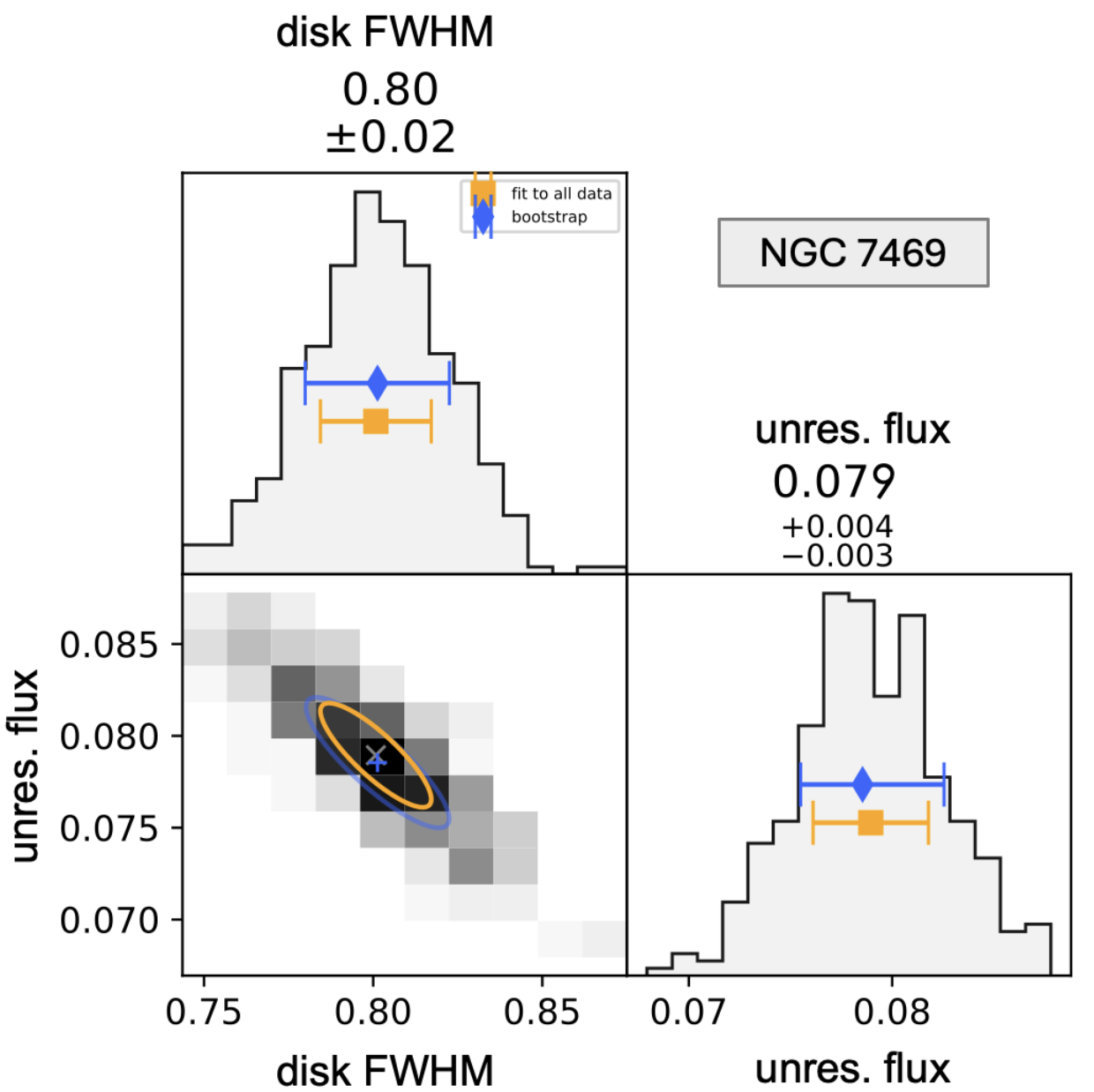}
    \caption{Corner plot of the single-source fit for NGC 7469, showing marginalized posteriors and parameter correlations from bootstrap resampling. The diagonal panels show the central FWHM and unresolved flux fraction. Orange contours indicate the global-fit confidence intervals, while blue markers and contours show the bootstrap results.
}
\label{fig:gaussian}
\end{figure}
\begin{figure}[h!]
    \centering
    \includegraphics[width=1\linewidth]{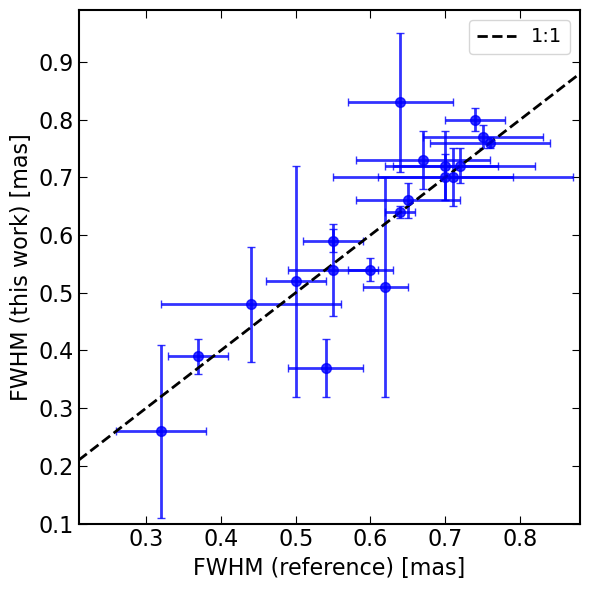}
    \caption{Comparison between the FWHM measurements obtained in this work and those reported in the literature. Error bars indicate the corresponding uncertainties, while the dashed line represents the one-to-one relation.}
    \label{fig:fwhm}
\end{figure}
The results obtained on different nights are consistent within $3\sigma$ for all objects, confirming the stability of the Gaussian central component.

An example of the posterior distribution obtained from the fit of the single central component for NGC~7469 is shown in Fig.~\ref{fig:gaussian}. All FWHM values are reported in Table~\ref{tab:fwhm} and are consistent with those reported in \cite{Cao24} and \cite{dexter_gravity20}, as shown in Fig.~\ref{fig:fwhm}.

\begin{table*}[t]
\centering
\caption{Angular FWHM sizes of the compact K-band central emission measured in this work and from previous studies. For each system, we report the number of detections above $5\sigma$, and the number of binary candidates retained after the validation procedure.}
\label{tab:fwhm}
\begin{tabular}{lccccccccc}
\hline\hline
Source &  K (mag) &FWHM (ref) [mas] & FWHM (this work) [mas] & $>5\sigma$ detections & Binary candidate\\
\hline
Ark 120     &10.8&    $0.70 \pm 0.09$ & $ 0.70\pm 0.04$&2&0 \\
HE 1029$-$1401  &12.2 & $0.64 \pm 0.07$ & $ 0.83\pm 0.12 $& 45&0 \\
IC 4329A       &10.8 & $0.64 \pm 0.02$ & $ 0.64\pm 0.01$& 0&0\\
IRAS 09149--6206 &9.8 & $ 0.54\pm 0.05$ & $0.37\pm0.05$& 0&0\\ 
IRAS 13349+2438 &10.1 & $0.76 \pm 0.08$ & $ 0.76\pm 0.01$& 4&0\\
Mrk 1239        &9.9 & $0.55 \pm 0.04$ & $ 0.59 \pm 0.02 $& 0&0\\
Mrk 335         &10.5&$0.44 \pm 0.12$ &$0.48 \pm 0.10$ & 0 &0\\

Mrk 509        &11.0  & $0.62 \pm 0.03$ & $ 0.51\pm 0.19$& 0 &0\\
NGC 3227       &11.0&   $0.50 \pm 0.04$ & $ 0.52\pm 0.20$& 0&0\\
NGC 3783        &9.8 & $0.72 \pm 0.10$ & $ 0.72\pm 0.03$&2 &1\\ 

NGC 4593       &11.6&   $0.37 \pm 0.04$ & $ 0.39\pm 0.03$&2 &0\\
NGC 7469       &11.1&   $0.74 \pm 0.04$ & $ 0.80\pm 0.02 $&0 &0\\
NGC 7603       &10.8&   $0.67 \pm 0.09$ & $ 0.73 \pm0.05 $&12 &0\\
NGC 1365       &10.0&   $0.55 \pm 0.06$ & $ 0.54 \pm 0.08 $&0 &0\\ 
PGC 89171      &11.1&   $0.65 \pm 0.07$ & $ 0.66\pm 0.03 $& 4&0\\
PDS 456        &11.0&   $0.60 \pm 0.03$ & $ 0.54\pm 0.02$&0 &0\\
UGC 545        &10.7&   $0.70 \pm 0.07$ & $ 0.72\pm 0.06  $&0&0\\
UGC 11763      &11.4&   $0.75 \pm 0.08$ & $ 0.77\pm0.02 $& 6 &0\\
3C120      &11.6  & $0.71 \pm 0.16$ & $0.70 \pm 0.05$ & 0&0 \\ 
3C273      &10.0  & $0.32 \pm 0.06$ & $ 0.26\pm0.15 $& 0 &0\\ 

\hline
\end{tabular}
\end{table*}

\subsection{Step 2: Companion search}
After deriving the best-fit model for the central source in Step 1, we fixed its parameters and searched for an additional unresolved component in the field of view. Candidate companion positions were explored through a two-dimensional grid search, adding an unresolved point source to the best-fit central-source model and fitting the interferometric observables at each grid position.
Here, the position of the first component is used only to define the coordinate origin, while the position and flux of the secondary component are allowed to vary freely. Thus, fixing the first component at the origin does not impose that the origin coincides with the centre of the total K-band light distribution.

We explored offsets between $-40$ and $+40$~mas in both right ascension and declination, with a step size of 2~mas (approximately $\lambda/B_{\rm max}$). The best-fit solution was identified from the global minimum of the reduced $\chi^2$ map, while parameter uncertainties were estimated through bootstrapping. For each grid position, we computed the detection significance and the companion-to-primary flux ratio. 
Candidate companions were initially identified as local maxima with a detection significance exceeding the adopted threshold of $5\sigma$. The detection significance reported by the grid search should not be interpreted as a global false-alarm probability. The grid comprises thousands of positions, and the corresponding significance values are not statistically independent because of the interferometric response and the finite $uv$ coverage. As a result, multiple local maxima can exceed the nominal $5\sigma$ threshold even in the absence of a physical companion. This can arise from the structured response of the interferometric sampling and from deviations from the assumptions underlying the formal significance estimate, rather than simply from underestimated measurement uncertainties. 
We therefore adopt the $5\sigma$ threshold as an initial candidate-selection criterion, while the statistical and instrumental reliability of each candidate is assessed through the subsequent validation procedure. As a consistency check, we also repeated the search using a $3\sigma$ threshold. This resulted in additional marginal candidate positions for some targets, with up to 50 candidate positions for individual targets; however, none of these additional candidates passed the subsequent injection--recovery and PSF validation steps. The final validated candidate sample is therefore unchanged. The number of initial candidate positions can depend on the specific $uv$ coverage of each target, as sparse or anisotropic sampling can produce a more structured interferometric response. However, the subsequent injection--recovery and PSF validation steps explicitly incorporate the $uv$ sampling of each target and assess the reliability of individual candidates.

The first step of the validation procedure is illustrated in the first panel of Fig.~\ref{fig:workflow}. The flux map obtained from the grid search is color-coded according to the detection significance. For visualization purposes only, to highlight the residual structure and possible additional components, the best-fit central Gaussian component is subtracted from the model, and its position is marked by a yellow star. Candidate companions are indicated by green circles.

Table~\ref{tab:fwhm} lists the initial $>5\sigma$ candidates identified for each source. These candidates were subsequently validated following the workflow shown in Fig.~\ref{fig:workflow}. The validation procedure, described in the following subsections, combines injection--recovery simulations with PSF-based tests to reject spurious signals, retaining only robust companions. Sources with no surviving candidates therefore show no evidence for an additional compact companion beyond the central Gaussian component.

\subsubsection{Injection--recovery validation}
\label{sec:injrec}

Interferometric companion searches are affected by position-dependent sensitivity variations arising from the $uv$ coverage, the noise properties of the observables, and the coupling between the binary parameters. Therefore, the detection of a companion at a given location requires a quantitative estimate of the local detection limit. As the first validation step of the workflow (Fig.~\ref{fig:workflow}), we quantified the sensitivity of the \grav data using injection--recovery simulations performed with the {\tt PMOIRED} framework.
Injection--recovery simulations are widely used to characterize source-detection completeness and selection effects in astronomical surveys (e.g., \cite{Leethochawalit22}). Here, we adapt this approach to interferometric data to quantify the local detection sensitivity as a function of position across the field of view.

For each simulation, an artificial companion with a prescribed flux ratio and position was injected directly into the original interferometric data. The resulting data were then analysed using the same binary model and fitting procedure adopted in the companion search. We therefore did not remove or subtract previously identified candidate companions before performing the injections; any real source present in the original data remains part of the simulated dataset. The detectability of the injected companion was assessed by comparing the $\chi^2$ of the best-fitting single-source and binary models. We adopted the significance prescription implemented in {\tt PMOIRED}, which converts the improvement in $\chi^2$ between the two models, accounting for the three additional free parameters of the companion (flux ratio and two positional coordinates), into an equivalent Gaussian significance. This use of $\Delta\chi^2$ is only for assessing the significance of the additional component and is independent of the bootstrap-derived parameter uncertainties described in Sect.~\ref{sec:fwhm}.

By injecting companions with flux ratios between 0.01 and 1 over the full field of view, we constructed two-dimensional detection-limit and detection-probability maps. At each position, the injected companion was considered recovered when its inclusion produced a detection significance of at least $5\sigma$. 
The detection significance is computed from the improvement in $\chi^2$ between the best-fit model without the companion and the model including the companion, accounting for the three additional free parameters of the companion (flux and position). The detection-limit maps provide the minimum detectable flux ratio at each position, while the probability maps quantify the fraction of injected companions recovered according to this criterion.

The second panel of Figure~\ref{fig:workflow} shows the resulting detection-limit map, with the color scale representing the minimum detectable flux ratio across the field of view. Lower values indicate regions of higher sensitivity. The candidate companions identified in the grid search are overlaid as open circles, with their color encoding their measured flux ratios, while the central source is marked by the gold star at $(\Delta\mathrm{RA},\Delta\mathrm{Dec})=(0,0)$. Comparing the candidate flux ratios with the local detection limit provides a visual representation of the sensitivity to each candidate.
Candidates whose measured flux ratios fell below the local detection limit were rejected, whereas those exceeding it proceeded to the subsequent PSF validation. Because the injections are performed on the original data, this procedure also accounts empirically for the effect of any pre-existing compact secondary emission on the recovery of an additional component.

\subsubsection{PSF validation}
\label{sec:psfvalidation}

The second validation step aims to distinguish genuine astrophysical companions from artifacts produced by the interferometric point spread function (PSF). Only candidates that passed both the significance selection and the injection--recovery validation were subjected to this analysis.

The sparse and anisotropic $uv$ coverage of long-baseline interferometry produces a structured dirty beam with positive and negative sidelobes. These sidelobes are not necessarily radially symmetric and can have different amplitudes at different positions and directions from the central peak, potentially mimicking faint companions. To characterize the PSF structure specific to each observation, we constructed a two-dimensional interferometric dirty beam directly from the observed $uv$ coverage. We collected all unflagged $(u,v)$ points, added their Hermitian counterparts $(-u,-v)$, gridded the resulting sampling function, and computed its inverse Fourier transform. The real part of the resulting dirty beam was then normalized to its peak intensity.

To quantify the strength of the sidelobes, we excluded only the immediate central PSF peak using a fixed $21\times21$ pixel mask, corresponding to 10 pixels on either side of the peak along each image axis. This mask is a purely geometric exclusion region and is not defined from the PSF FWHM; its purpose is to prevent the central PSF core from setting the sidelobe threshold. The sidelobe map was then defined as the absolute value of the normalized PSF outside this central region, thereby accounting for both positive and negative sidelobes.

We defined the empirical strong-sidelobe threshold as 50\% of the maximum absolute sidelobe amplitude measured outside the central $21\times21$ pixel region:
$
T_{\rm sidelobe}
=
0.5 \times
\max_{\rm outside\ core}
\left|\mathrm{PSF}\right|.
$
For each candidate, its angular offset from the central AGN was converted to the corresponding coordinates on the PSF grid, and the local PSF response was evaluated as $
R_{\rm PSF}
=
\left|\mathrm{PSF}(x_{\rm cand},y_{\rm cand})\right|.
$
A candidate was rejected as a likely PSF artifact if
$
R_{\rm PSF} > T_{\rm sidelobe} + 3\sigma_{\rm noise}.
$
Here, $\sigma_{\rm noise}$ represents the standard deviation of the noise in the PSF response. This test therefore depends only on the local PSF response at the candidate position. It does not impose a flux-ratio threshold: the sensitivity to the companion flux ratio is determined independently by the injection--recovery analysis.

The third panel of the workflow shown in Figure~\ref{fig:workflow} illustrates this validation procedure. The color scale shows the normalized dirty beam, with the central AGN position, corresponding to the PSF peak, marked by a star. The cyan contours delimit the regions where the absolute PSF response exceeds 50\% of the maximum sidelobe amplitude measured outside the central $21\times21$ pixel region. Candidate positions are overlaid on the PSF. Candidates located within these strong-sidelobe regions are flagged as potential PSF artifacts, whereas candidates outside these regions are not rejected by this criterion.

Only candidates that passed the PSF-artifact check, together with the preceding significance and injection--recovery criteria, were retained as robust detections.

\section{Binary Candidates}
\label{sec:binarycandidates}
In this section, we present the results of the companion-search analysis for sources exhibiting at least one candidate detection above $5\sigma$ in the sample of 20 targets (see Table~\ref{tab:fwhm}). Candidate robustness is assessed through a combination of injection--recovery tests, local flux-ratio detection limits, PSF contamination diagnostics, and morphological consistency checks across independent image reconstructions.

The companion search identified several candidate secondary components across the sample. However, only one candidate satisfies all the criteria required for robust binary detection. As described in Sect.~\ref{sec:injrec}, the initial $>5\sigma$ threshold therefore represents only the first stage of the selection and does not, by itself, imply the presence of a physical companion. While GRAVITY can achieve high sensitivity on relatively short timescales, more extensive $uv$ coverage may be required to robustly distinguish genuine faint sub-structures from interferometric artifacts. Future observations with complementary baseline configurations may therefore help to further assess the nature of individual candidates.

\subsection{Rejected candidates}
\label{sec:no_binary_detection}

Below, we discuss the sources for which the candidate signals do not meet all the criteria for a robust binary detection. As described in Sect.~\ref{sec:analysis}, candidates are considered robust only if they (i) exceed the adopted $5\sigma$ significance threshold, (ii) have a measured flux ratio above the local detection limit established by the injection--recovery simulations, and (iii) are not associated with strong PSF sidelobes or other point-symmetric interferometric features. Candidates failing any of these criteria are rejected.

\paragraph{Candidates associated with PSF structure.}
In Ark~120, two approximately point-symmetric secondary features are identified at offsets of roughly $(+1,+40)$~mas and $(-1,-40)$~mas, with comparable flux ratios of $\mathrm{FR}\simeq5\%$. Although one of these features formally exceeds the local $5\sigma$ significance threshold, both candidates are aligned with the principal elongation axis of the interferometric response and exhibit a highly symmetric morphology. The PSF diagnostics indicate that this pattern is consistent with the anisotropic $uv$ coverage and the resulting sidelobe structure. The candidates are therefore rejected as PSF-related artefacts rather than physical companions.

The absence of a robust continuum-scale companion in Ark~120 is also consistent with previous studies of this source. Ark~120 has been proposed as a candidate sub-parsec SMBH binary based on long-term optical variability and a possible $\sim20$~yr periodicity \citep{Li17}. However, for such compact systems, the expected binary separation is well below the direct continuum imaging capabilities of the interferometric observations, while BLR spectro-interferometric measurements provide a more sensitive diagnostic through differential phase signatures produced by the orbital motion of two BLRs \citep{Kovacevic20,Songsheng19}. Moreover, multi-wavelength monitoring of Ark~120 shows a tight correlation between the X-ray, UV, and optical emission, with inter-band lags following the expected $\tau\propto\lambda^{4/3}$ relation, favouring a standard accretion-disc reprocessing scenario and providing no evidence for additional orbital-scale variability \citep{Lobban20}. Thus, neither the continuum reconstruction presented here nor the available multi-wavelength constraints provide independent evidence for a compact binary in Ark~120.

A similar behaviour is observed in NGC~4593, where two secondary features with nearly identical flux ratios of $\mathrm{FR}\simeq2.2\%$ are aligned with the principal elongation axis of the synthesized beam. Neither candidate exceeds the adopted $5\sigma$ detection threshold. Lowering the threshold to $3\sigma$ reveals the same elongated pattern, indicating that these features are associated with the underlying interferometric response rather than with a distinct compact source. The PSF analysis therefore provides no evidence for a genuine asymmetric compact component beyond the central Gaussian nuclear component, and the source remains consistent with a single nuclear component within the sensitivity limits of the observations.

NGC~7603 represents a further example of PSF-related candidate features. The initial grid search identifies 12 candidate secondary components, all formally exceeding the $5\sigma$ significance threshold but none reaching $8\sigma$ (see Fig.~\ref{fig:workflow}). Most candidates are located within regions of strong PSF sidelobe response. The PSF diagnostics therefore identify these features as consistent with the interferometric response rather than as physically distinct sources. None of the candidates survives the full validation procedure.

The rejection of these candidates is particularly relevant given the complex environment of NGC~7603. The galaxy is interacting with the nearby companion PGC~71041 and exhibits tidal distortions and a luminous filament connecting the two galaxies \citep{Tohline76,LopezCorredoira02}. Its nucleus is radio weak and core dominated, with no evidence for a powerful relativistic jet or resolved jet structure on milliarcsec scales \citep{Condon98,Becker95,Lal04}. NGC~7603 is also a changing-look AGN, having transitioned between type~1, type~2, and intermediate spectral states over decadal timescales \citep{Kollatschny00,Trippe10,Jana+25}. Although this complex nuclear environment could in principle generate compact asymmetric emission on $\sim20$~pc scales through interaction-induced nuclear structures, gas inflows, or AGN-driven outflows, the interferometric data do not provide evidence for a physical secondary component. We therefore find no robust evidence for a binary AGN in NGC~7603.

\paragraph{Candidates below the empirical detection limits.}

A second group comprises sources in which the candidate features identified in the initial search do not satisfy the local flux-ratio detection limits established by the injection--recovery simulations. This is the case for HE~1029--1401, IRAS~13349+2438, PGC~89171, and UGC~11763.

In HE~1029--1401 and IRAS~13349+2438, the initial search identifies multiple secondary features across the interferometric field, including several candidates that formally reach the adopted $5\sigma$ significance threshold. However, their measured flux ratios lie below the local detection limits derived from the injection--recovery maps. These candidates are therefore not considered reliably detectable at their respective positions. The remaining features do not provide a robust signal in the subsequent PSF validation. The observed complexity is consequently attributed to low-level fluctuations and limited $uv$ coverage rather than to genuine compact companions. Within the sensitivity limits of the present observations, both sources are consistent with the absence of an additional companion component beyond the central Gaussian nuclear component.

UGC~11763 provides a similar example. The initial search identifies multiple candidate secondary components, including features formally exceeding the $5\sigma$ threshold. However, their flux ratios are at or below the local detection limits derived from the injection--recovery simulations, with $\mathrm{FR}\lesssim0.1$, indicating that they are not robustly detectable given the local sensitivity of the data. PSF diagnostics further reduce the number of viable candidates, leaving only a residual feature at approximately $(x,y)\simeq(11,-7)$~mas. Although this feature does not appear to be directly associated with a prominent PSF sidelobe, its flux ratio remains below the local detection limit and it is therefore not considered a robust detection. The nuclear emission of UGC~11763 is consequently consistent with a single compact component, in agreement with its known disc--corona dominated emission and multi-phase circumnuclear absorbing material \citep{Cardaci09,Jiang22}.

PGC~89171 also shows several low-significance secondary features, with flux ratios of approximately $\mathrm{FR}\sim2.5$--$3.8\%$. These candidates do not remain above the local detection limits after the injection--recovery validation, and no persistent compact asymmetry is recovered. The global morphology is instead consistent with a moderately inclined, disk-like nuclear structure with an ellipticity of $\sim0.46$. The residual features are therefore more naturally interpreted as low-level substructure within a mildly asymmetric or clumpy nuclear distribution rather than as distinct compact companions.

Overall, the candidate features identified in the initial significance maps do not provide evidence for confirmed binary AGN. Candidates are rejected either because their spatial distribution and morphology are consistent with PSF sidelobes or point-symmetric interferometric artefacts, or because their flux ratios fall below the empirical detection limits established through injection--recovery tests. Within the sensitivity limits of the present observations, all sources discussed in this section are therefore classified as non-detections.

\subsection{Binary Candidate}
\label{sec:binary_candidates}

While the sources discussed in Sect.~\ref{sec:no_binary_detection} do not show any robust evidence for a secondary compact component after the full validation procedure, NGC~3783 represents a different case. In this source, one candidate survives the PSF filtering and reconstruction-consistency checks, and the interferometric data reveal a persistent departure from centro-symmetric emission, with a closure phase that is consistently different from zero (bottom panel in Fig.~\ref{fig:cp}). Although the available observations do not allow us to establish whether this asymmetry is produced by a genuine secondary nucleus, NGC~3783 is therefore retained as a binary candidate.

NGC~3783 is a nearby Seyfert~1 galaxy at $z=0.0097$, hosting a well-studied active nucleus with a prominent multi-phase warm absorber. High-resolution X-ray spectroscopy reveals a stratified ionized outflow with multiple components spanning a wide range of ionization states and velocities \citep{Kaspi01,Reeves04,Mehdipour17}. Integral-field spectroscopy further supports a complex nuclear environment composed of both rotating and outflowing gas components \citep{MullerSanchez11}, while previous mid-infrared interferometric observations with the VLTI/MIDI instrument revealed a polar-elongated dust component on parsec and sub-parsec scales \citep{Honig+13,Burtscher13,LopezGonzaga16}. More recently, \citet{gravity_taro_NGC3783} reported a rotating broad-line region and an asymmetric hot-dust structure on sub-parsec scales, further supporting the presence of a complex and non-axisymmetric nuclear environment.

\begin{figure}[t]
    \centering
    \includegraphics[width=1\linewidth]{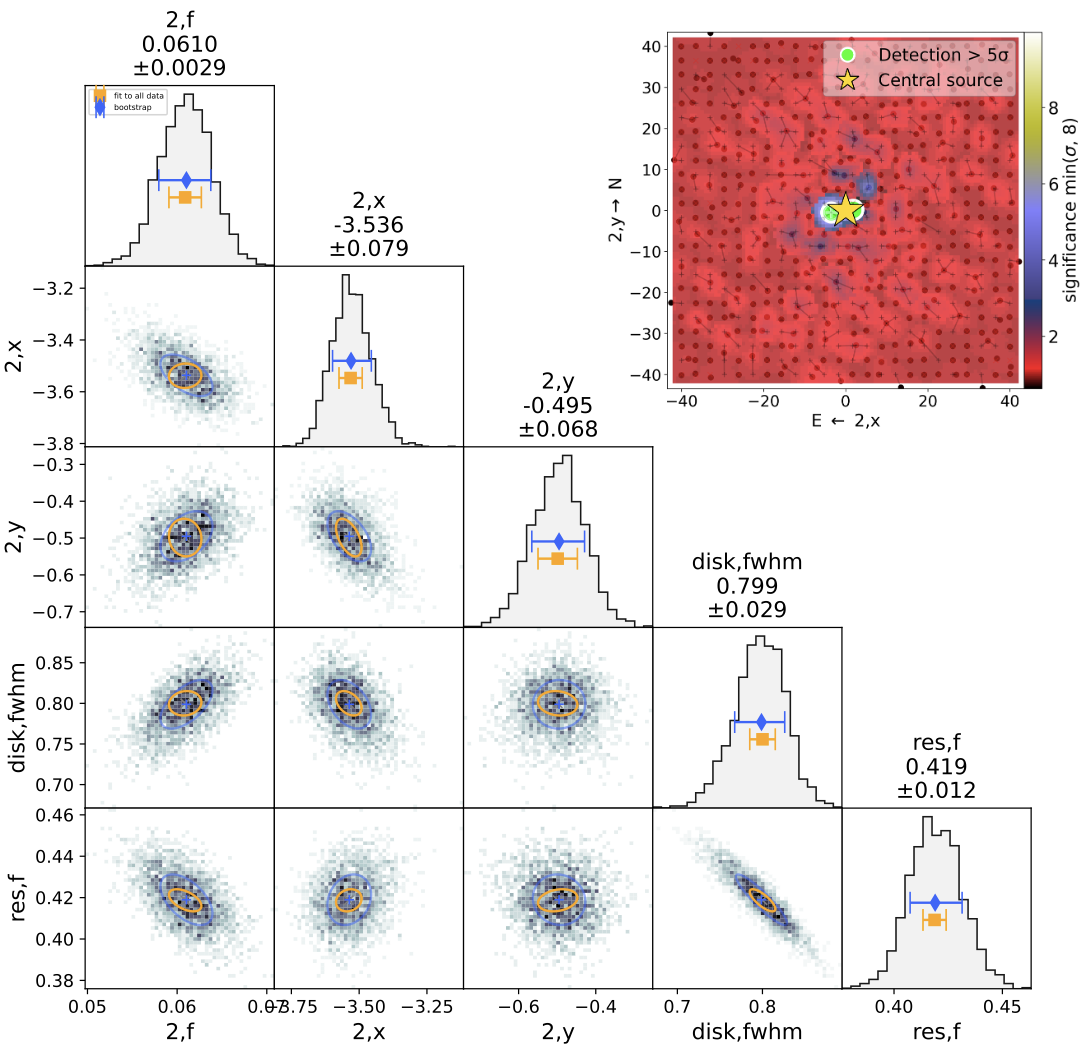}
    \caption{Posterior distributions and parameter correlations for the fit with a secondary component in NGC 3783, showing its flux fraction, position $(x,y)$, disk FWHM, and unresolved flux fraction. Blue and orange markers indicate the best-fit and bootstrap estimates, respectively. The upper-right panel shows the detection-significance map, with the central source marked by a yellow star and $>5\sigma$ detections highlighted in green.}
    \label{fig:gaussngc3783}
\end{figure}

In our analysis, two secondary candidates are initially identified in the initial reconstruction of NGC~3783, as shown in the upper-right panel of Fig.~\ref{fig:gaussngc3783}. Following PSF filtering and reconstruction-consistency checks, one of the two features is not recovered and is therefore attributed to reconstruction or PSF-related artefacts.
The remaining candidate is identified at $(x,y)\simeq(-3.5,-0.5)$~mas in the grid search and is recovered at a consistent position in the subsequent best-fit two-source model. This model consists of a central Gaussian disk with an optimized FWHM and a resolved background flux component, together with an additional point source representing the candidate companion. The candidate has a flux ratio of $\mathrm{FR}\simeq0.06$. 
Such a flux ratio is physically plausible for a binary AGN, as the luminosities of the two AGN can differ substantially owing to differences in their instantaneous accretion rates. In particular, the flux ratio does not directly trace the black-hole mass ratio, and large luminosity contrasts are found both in simulations and observations of merger AGN \citep{Volonteri22,Perna2025}.
The posterior distributions of this fit are shown in the lower panels of Fig.~\ref{fig:gaussngc3783}, while the corresponding fits to the T3PHI and $V^2$ data are shown in Fig.~\ref{fig:fitcpv_NGC3783} in Appendix~\ref{app:binaryfit}.

Given the known inclination of the nuclear dust structure in NGC~3783 \citep{gravity_taro_NGC3783}, an elliptical Gaussian may provide a more physically motivated phenomenological description of the central emission than a circular Gaussian. 
We therefore tested whether the evidence for a secondary component depends on the assumed geometry of the central emission.\\ 
We first compare the models with and without a secondary component for both central brightness distributions. 
For the circular model, adding the point-like secondary decreases $\chi^2$ from $258.52$ to $213.00$, corresponding to $\Delta\chi^2=-45.53$ for three additional free parameters, while the BIC decreases from $275.6$ to $255.7$. Similarly, for the elliptical model, adding the secondary decreases $\chi^2$ from $258.32$ to $212.09$, corresponding to $\Delta\chi^2=-46.22$ for three additional free parameters, while the BIC decreases from $291.7$ to $271.8$. Thus, the secondary component is strongly preferred over the corresponding single-component model for both central brightness distributions. This demonstrates that the improvement in the fit is not simply a consequence of allowing the central component to deviate from circular symmetry.\\
We then repeated the fit using an elliptical Gaussian for the central component, allowing both its inclination and position angle to vary while retaining the point-like secondary component. The elliptical model yields $\chi^2=212.09$ for 5059 degrees of freedom, compared with $\chi^2=213.00$ for 5061 degrees of freedom in the circular model. Thus, introducing two additional free parameters improves the fit by only $\Delta\chi^2=-0.90$, while the BIC increases from $255.7$ to $271.8$. The elliptical model is therefore not statistically favored over the circular model according to the BIC. Importantly, the inferred properties of the secondary component are essentially unchanged. The flux ratio changes from $F_2/F_1=0.061\pm0.003$ to $0.063\pm0.002$, while its position changes from $(-3.54\pm0.08,-0.50\pm0.07)$~mas to $(-3.60\pm0.05,-0.56\pm0.05)$~mas with position angles of $262^\circ$ and $261^\circ$. Thus, allowing the central component to be elliptical does not affect the inferred position or flux ratio of the candidate secondary.

However, the complex nuclear environment of NGC~3783 provides several possible physical explanations for the observed signal. In particular, the polar dust emission and the asymmetric hot-dust structure reported on parsec and sub-parsec scales could naturally produce localized non-axisymmetric emission. Conversely, a compact secondary nucleus cannot be excluded by the present data.
As discussed by \citet{gravity_taro_NGC3783}, the observed $\sim5$ mas separation corresponds to a projected distance of $\sim0.6$ pc, which represents a lower limit to the true separation. However, considering the inferred geometry of the nuclear region, with a torus opening angle of $\sim45^\circ$ and a line-of-sight inclination of $\sim23^\circ$, the projection correction remains modest, and does not qualitatively affect the interpretation. 
In the binary scenario, adopting a primary SMBH mass of $\log(M_{\rm BH}/M_\odot)=7.68$ \citep{gravity_taro_NGC3783}, the expected orbital velocity at this distance would be $\sim900$ km s$^{-1}$, while the corresponding orbital period would be several thousand years, making dynamical or periodic signatures inaccessible on observational timescales. 
Conversely, the observed $K$-band emission can be reproduced by a dust cloud externally heated by the primary AGN, requiring a gas mass of only $\sim20,M_\odot$ and a density of order $\sim10^4$ cm$^{-3}$ \citep{gravity_taro_NGC3783}. Such values remain physically plausible for circumnuclear gas. We therefore retain NGC~3783 as a binary candidate, but do not consider the present evidence sufficient for a robust binary SMBH detection.

More generally, for a given AGN bolometric luminosity, redshift, and secondary-to-primary flux ratio, we define a critical angular separation, $\theta_{\rm crit}$, beyond which an AGN-heated gas cloud would have to be too massive and dense to remain consistent with the adopted unresolved-cloud scenario. Here, $\theta_{\rm crit}$ is the maximum projected separation at which a gas cloud, assumed to have a critical density of $n_{\rm H}=10^7$ cm$^{-3}$ and to remain unresolved by GRAVITY, can reproduce the observed secondary emission. Thus, separations smaller than $\theta_{\rm crit}$ allow an AGN-heated gas cloud to remain a viable explanation, whereas at larger separations the required cloud properties become increasingly difficult to reconcile with a compact gas clump.

\begin{figure}
    \centering
    \includegraphics[width=1\linewidth]{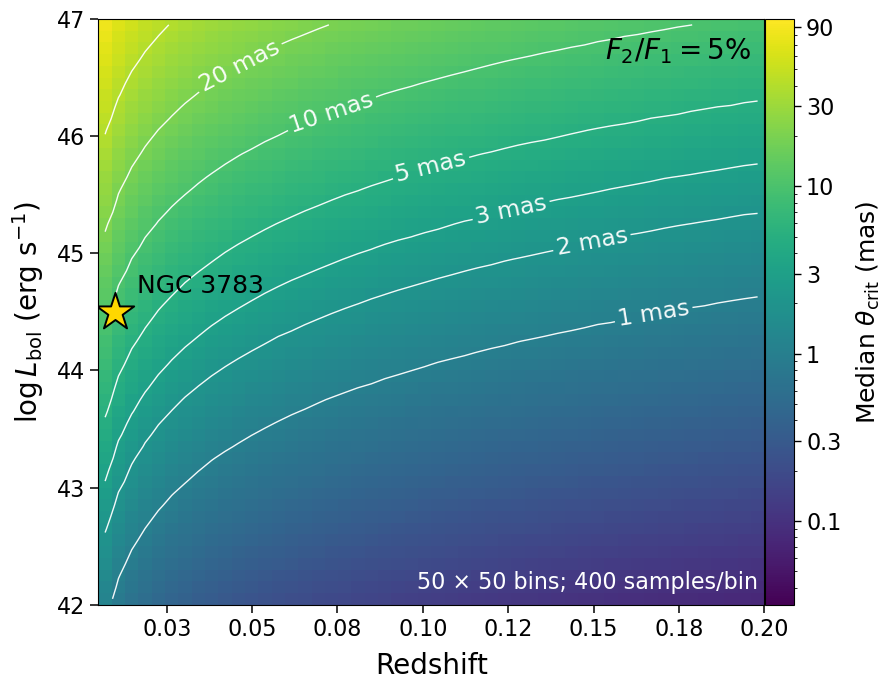}
    \caption{Critical angular separation for an AGN-heated gas-cloud interpretation. The color scale and contours show the median $\theta_{\rm crit}$ as a function of AGN $L_{Bol}$ and redshift, assuming a $5\%$ secondary-to-primary $K$-band flux ratio, $n_{\rm H}=10^7,{\rm cm}^{-3}$, and a maximum cloud diameter of 3 mas. The star marks NGC~3783, whose $\sim5$mas separation lies below the corresponding $\theta_{\rm crit}$.}
    \label{fig:MC_crit_sep}
\end{figure}

We adopt a maximum cloud diameter of 3 mas, corresponding to the maximum angular size of a cloud that remains unresolved by GRAVITY, and a critical gas density of $n_{\rm H}=10^7$ cm$^{-3}$. These assumptions set the maximum emitting volume and gas density of the cloud and therefore determine the maximum $K$-band emission that can be produced by an AGN-heated cloud at a given distance. \\
The resulting critical separation is shown as a function of luminosity and redshift in Fig.~\ref{fig:MC_crit_sep}. NGC~3783, with a separation of $\sim5$ mas, lies approximately a factor of two below the corresponding critical separation, indicating that an AGN-heated gas cloud remains a viable explanation for its compact secondary emission. Conversely, systems with separations larger than $\theta_{\rm crit}$ would require a cloud with properties different from those adopted here to produce sufficient $K$-band emission, making the clump interpretation increasingly difficult. Such systems would therefore provide stronger evidence that the compact secondary is associated with a second AGN rather than an AGN-heated gas cloud.

\section{Binary AGN fraction}
\label{sec:binary_fraction}

Our sample of 20 bright AGN at $z<0.25$, selected with a uniform near-infrared sensitivity threshold and optimized for observations with \grav \citep{Gravity+17_1}, enables a systematic search for compact secondary emission components on parsec scales. The observations probe projected separations of approximately $5$--$40$ mas, corresponding to sub-pc to tens-of-pc scales depending on source distance, with the lower limit set by the interferometric angular resolution and the upper limit by the VLTI single-mode field of view.

Among the 20 targets, one source (NGC~3783) exhibits significant departures from centro-symmetric emission, manifested through non-zero closure phases and improved fits when an additional compact component is included alongside a central Gaussian emission profile. These asymmetric signals are independently recovered across reductions and remain qualitatively stable under PSF diagnostics and image reconstruction tests.

However, the presence of a compact asymmetric component does not uniquely imply a binary AGN. Previous interferometric studies have demonstrated that compact dusty clouds, polar emission, and outflow-related structures can produce offset or localized emission components capable of generating similar closure-phase signatures (e.g. \citealt{gravity_taro_NGC3783}). In particular, the observed asymmetry in NGC~3783 is plausibly consistent with the polar dust emission previously identified on parsec scales.

We therefore identify one compact asymmetric component among the 20 targets, corresponding to an observed incidence of $1/20=5\%$. If this candidate is interpreted as a genuine binary AGN detection, the Poissonian one-sided $1\sigma$ upper limit on the mean number of detectable binaries is $<3.3$, corresponding to $f_{\rm bin}<16.5\%$. However, given that the observed asymmetry in NGC~3783 can also be explained by asymmetric circumnuclear dust emission, we do not regard it as a confirmed binary SMBH detection. Under this more conservative interpretation, the number of confirmed binary detections is zero, yielding a Poissonian one-sided $1\sigma$ upper limit of $<1.8$ systems, or $f_{\rm bin}<9.2\%$.
The present sample therefore provides an empirical constraint on the incidence of detectable parsec-scale binary AGN among bright AGN at $z<0.25$ with $K<12.5$. The derived limit should be regarded as an upper limit on the fraction of detectable binary AGN configurations, rather than as a measurement of the intrinsic binary fraction.

The completeness of the search was characterized through injection--recovery tests performed independently for each target (see Sect.~\ref{sec:injrec}). For each dataset, artificial unresolved secondary components were injected at a grid of positions across the field of view and for a range of flux ratios. At each position, we determined the minimum flux ratio for which the injected companion was recovered according to the adopted detection criterion. Because the injections are performed directly on the observed interferometric datasets, the resulting recovery fractions naturally account for the actual $(u,v)$ coverage, data quality, and position-dependent sensitivity of each observation. On average, companions with flux ratios $\mathrm{FR}>0.1$ are recovered in $\sim95\%$ of the tested configurations, indicating that the survey is highly complete to companions above this contrast level within the probed separation range.

We adopt $\mathrm{FR}>0.1$ as the representative contrast threshold for the full sample because this level provides a high and relatively homogeneous completeness across the targets. Lower flux ratios, such as $\mathrm{FR}=0.05$, can be reached for individual objects with particularly extensive observations and favourable $(u,v)$ coverage, as is the case for NGC~3783, but this sensitivity is not achieved uniformly across the full sample. We therefore do not adopt $\mathrm{FR}=0.05$ as a common threshold for the survey. The resulting limits define the regions of parameter space in which a companion would have been detectable with the available \grav data. Conversely, a non-detection does not exclude companions below the local sensitivity limit, and the constraints on the binary fraction should therefore be interpreted as applying primarily to companions above the corresponding detection threshold. Binary systems with companions below the \grav sensitivity limit or outside the probed angular scales therefore remain unconstrained by the present survey.

\section{The Horizon-AGN simulation}
\label{HAGN}

\hagn~\citep{Dubois14} is a hydrodynamical cosmological simulation with a volume of $(142,{\rm cMpc})^3$, assuming a standard $\Lambda$CDM cosmology. The spatial resolution is $\Delta x=1,{\rm kpc}$, and SMBHs are seeded with an initial mass of $10^5,\msun$. SMBHs are not artificially pinned to galaxy centres, and their evolution includes dynamical friction from gas and stars \citep{Dubois2013}. Galaxies and dark matter haloes are identified using the \textsc{AdaptaHOP} finder \citep{aubertetal04}, with their centres determined using the shrinking-sphere method \citep{poweretal03}.

We use the binary AGN catalog generated from post-processed \hagn~data \citep{2020MNRAS.498.2219V, Hippolyte2025}. In \hagn, each SMBH is assigned a unique ID that is tracked until it numerically merges with another SMBH when their separation falls below $4\Delta x$. Here, $\Delta x=1$~kpc is the spatial resolution of the simulation, whereas $4\Delta x=4$~kpc is the numerical SMBH-merger criterion and should not be interpreted as the spatial resolution. The merged SMBH retains the ID of the primary object, while the secondary ID is discarded. Numerical mergers therefore occur at separations of $\sim4$~kpc, well above the scales probed by \grav~ at $z<0.5$. Mergers are identified at each simulation output ($\sim0.6{-}0.7$~Myr) by tracking disappearing SMBH IDs. Only mergers occurring within $\max(2R_{\rm eff},4\Delta x)$ of galaxy centres are included in the catalog, where $R_{\rm eff}$ is the projected two-dimensional effective radius of the host galaxy, defined as the radius enclosing half of its stellar mass. This criterion restricts the catalog to SMBH mergers occurring within the central regions of their host galaxies.

In \hagn~, numerical mergers occur when the separation of two MBHs is about $4,{\rm kpc}$, which is larger than the field of view of \grav~ in the low-redshift Universe ($z<0.5$). To make predictions for \grav~, we therefore evolve the MBH pairs in post-processing from the numerical-merger separation down to the innermost stable circular orbit (ISCO), adopting $r_{\rm ISCO}=6r_{\rm sch}$, where $r_{\rm sch}=2GM_{12}/c^2$ is the Schwarzschild radius corresponding to the total binary mass $M_{12}=M_1+M_2$.

The large-scale orbital decay is modelled through stellar dynamical friction following \citet{Li22}. We assume that each MBH is embedded in an isothermal stellar distribution and estimate the dynamical-friction timescale using a Chandrasekhar-type prescription,
\begin{equation}
t_{\rm df}=0.67
\left(\frac{d}{4{\rm kpc}}\right)^2
\left(\frac{\sigma_{\star}}{100\kms}\right)
\left(\frac{M_{\rm BH}}{10^8\msun}\right)^{-1}
\Lambda^{-1}{\rm Gyr},
\label{eq:tdf}
\end{equation}
where $M_{\rm BH}$ is the black-hole mass, $d$ is its distance from the galactic centre, and $\sigma_{\star}$ is the stellar velocity dispersion, approximated as $(0.25GM_{\star}/R_{\rm eff})^{1/2}$. Here, $M_{\star}$ is the total stellar mass of the host galaxy measured at the simulation output immediately preceding the numerical merger, and $\Lambda=\ln(1+M_{\star}/M_{\rm BH})$ is the Coulomb logarithm. The prescription includes a factor of $0.3$ to account for the non-circularity of typical MBH orbits. We consider only the stellar contribution to dynamical friction, which is expected to dominate over the corresponding contributions from gas and dark matter for the systems considered here \citep{2020ApJ...896..113L,2020ApJ...905..123L,2022MNRAS.510..531C}. For each pair, the dynamical-friction delay time is taken as the longer of the two individual timescales, typically corresponding to the secondary MBH.

The dynamical-friction evolution is followed until the pair reaches the influence radius,
\begin{equation}
r_{\rm inf}=R_{\rm eff}\left(\frac{4M_{12}}{M_{\rm gal}}\right),
\label{eq:rinf}
\end{equation}
where $M_{12}$ is the total binary mass and $M_{\rm gal}$ is the stellar mass of the host galaxy. At this stage, the two MBHs are assumed to form a gravitationally bound binary. The subsequent orbital decay from $r_{\rm inf}$ to $r_{\rm ISCO}$ is then modelled following \citet[][Sec.~2.3]{Li22}, including stellar hardening, viscous torques from a circumbinary disc, and gravitational-wave emission. The stellar density and velocity dispersion at $r_{\rm inf}$, corresponding to $\rho_i$ and $\sigma$ in \citet[][Eqs.~1--2]{Li22}, are extracted from the \hagn~ host galaxy at the end of the dynamical-friction evolution. This post-processing therefore connects the kiloparsec-scale evolution driven by dynamical friction to the subsequent bound-binary and gravitational-wave-driven phases.

To estimate the infrared detectability of the simulated binary AGN, we model the AGN emission from X-rays to the infrared using the mean AGN spectral energy distribution of \citet{2020MNRAS.495.3252S}. The bolometric luminosity is computed from the simulated accretion rate, averaged over $1,{\rm Myr}$ around the numerical-merger event, and the total binary mass. Dust attenuation is included by estimating the hydrogen column density within a sphere of radius $4,{\rm kpc}$, assuming a dust-to-gas ratio of $10^{-2}$ and a Milky Way extinction curve. The column density is given by
\begin{equation}
\log_{10}\left(\frac{N_{\rm H}}{\rm cm^{-2}}\right)=
22.54-0.23(1+z)
+
[0.65-0.1(1+z)]\log_{10}(f_{\rm Edd}).
\end{equation}
Single AGN are assigned $K$-band magnitudes using the analytical bolometric-to-infrared conversion of \citet{2020MNRAS.495.3252S}, while binary AGN are treated using the full SED-based model. This treatment assumes a spherically averaged distribution of obscuring material on 4~kpc scales and does not account for the anisotropic obscuration expected from a torus-like structure on parsec-to-tens-of-parsec scales. The predicted $K$-band luminosity, particularly that of the secondary SMBH, may therefore be overestimated for highly inclined systems or systems with intrinsically different dust distributions.
Consequently, the simulations do not capture possible orientation-dependent differences in the GRAVITY detectability of binary AGN. Such obscuration can modify the observed $K$-band flux ratio of the two components and therefore affect the fraction of simulated binaries satisfying the GRAVITY detectability criterion. Quantifying this effect would require incorporating an orientation-dependent torus model into the simulations, which is beyond the scope of the present analysis.

To connect the simulated binary population with \grav, we track the separation evolution $d_{12}$ and record the redshifts $z_{\rm FOV}$ and $z_{\rm RES}$ at which the projected separation becomes smaller than the instrument field of view and angular resolution, respectively. We adopt a field of view of $60$~mas and an angular resolution of $3$~mas for \grav.

\subsection{\grav-like observable binary AGN}
\label{binary AGN}

We identify binary AGN in the \hagn~catalog by requiring a total system Eddington ratio $f_{\rm Edd,sys}\geq0.01$, which we adopt as the minimum threshold for an SMBH binary to be considered actively accreting. We then restrict the sample to $z\leq0.5$, corresponding to the redshift range explored by the \grav~ analysis. We do not impose a lower redshift limit ($z<0.25$) in order to maximize the number of available systems, as no significant evolution is expected over this interval given the temporal resolution of the simulation \citep{Volonteri16}.

For each binary, \hagn~provides the masses of the two SMBHs at the numerical merger, when their separation reaches $4$~kpc, together with the total system Eddington ratio and the system $K$-band magnitude. The SMBH masses measured at $1$~kpc are used as fixed masses throughout the subsequent post-processed inspiral, since significant mass growth is not expected over the relatively short timescales associated with the evolution from kiloparsec to parsec separations. The system Eddington ratio and luminosity are retained as fixed constraints on the total emission, while the individual accretion states of the two SMBHs are not directly provided by \hagn.

\begin{figure}[]
\centering
\includegraphics[width=0.49\textwidth]{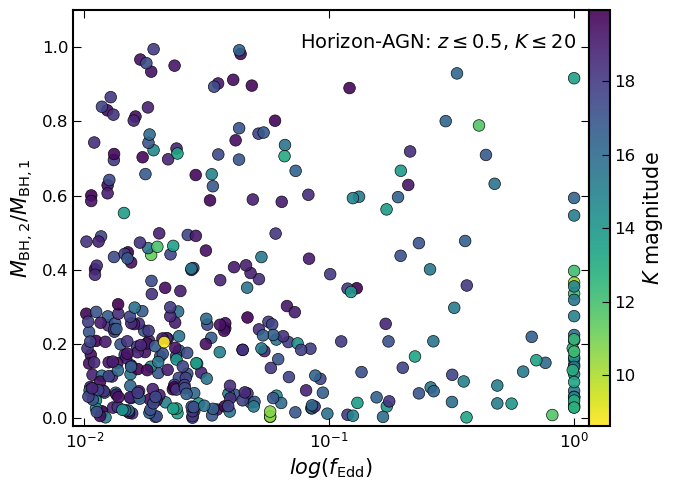}
\caption{Mass ratio of the two AGN at the numerical merger as a function of the total system Eddington ratio, $f_{\rm Edd,sys}$, for the Horizon-AGN binary AGN population at $z\leq0.5$ evolved to parsec-scale separations in post-processing. Points are colour-coded by $K$-band magnitude.}
\label{fig:sim_HAGN_0.5}
\end{figure}

We first investigate the intrinsic properties of the resulting low-redshift binary population before applying the \grav~ observability criteria. The primary SMBH is defined as the more massive component ($M_1\geq M_2$), and the mass ratio is expressed as $M_2/M_1$. Figure~\ref{fig:sim_HAGN_0.5} shows the mass-ratio distribution as a function of the total system Eddington ratio, with the systems colour-coded by their $K$-band magnitude.
Most systems have $f_{\rm Edd,sys}\leq0.1$, while only $\sim6\%$ reach values close to the Eddington limit (see Fig.~\ref{fig:sim_HAGN_0.5}). The SMBH accretion rate in \hagn~ is capped at $f_{\rm Edd}=1$. The separation distribution peaks below $\sim50$~pc, indicating that the population probes the late stages of SMBH pairing and merger evolution. The distributions of mass ratio, separation, and $K$-band magnitude reflect the combined effects of the intrinsic binary population and the post-processing procedure.

We then apply the \grav~ observability criteria to identify potentially detectable binary AGN. We first require the systems to lie within the angular separation range accessible to \grav. To reproduce the sensitivity of the observed \grav~ sample, we further require a total system magnitude of $K\leq12.5$ (Vega).
However, the total system luminosity alone is not sufficient to determine whether both SMBHs would be detected as separate components by \grav. In particular, the detectability of the secondary depends on the flux contrast $F_2/F_1$. This requires the luminosity of each SMBH individually, whereas \hagn~ provides only the total system Eddington ratio and does not track the individual accretion rates. We therefore use a post-processing Monte Carlo procedure to distribute the total accretion luminosity between the two SMBHs.

Starting from the system Eddington ratio provided by \hagn, we compute the total bolometric luminosity.
We then draw independent individual Eddington ratios, $f_{\rm Edd,1}$ and $f_{\rm Edd,2}$, from log-normal distributions motivated by observational constraints \citep{Schulze10,BrandonMerloni11}:
\begin{equation}
\log_{10}(f_{\rm Edd})
\sim
\mathcal{N}
(\mu_{\log f_{\rm Edd}},
\sigma_{\log f_{\rm Edd}}),
\end{equation}
adopting $\mu_{\log f_{\rm Edd}}=-1.0$ and $\sigma_{\log f_{\rm Edd}}=0.4$ dex as the fiducial values. This corresponds to a median Eddington ratio of $f_{\rm Edd}=0.1$, with approximately $68\%$ of the distribution lying in the range $0.04\lesssim f_{\rm Edd}\lesssim0.25$. These draws are used only to determine how the fixed total luminosity is partitioned between the two SMBHs and do not represent accretion rates independently predicted by \hagn.

For each Monte Carlo realization, the luminosities associated with the drawn Eddington ratios are rescaled by a common factor such that their sum exactly reproduces the total luminosity predicted by \hagn. Thus, the assumed distributions determine the relative accretion rates of the two SMBHs, and hence their luminosity contrast, while preserving the total luminosity of the simulated system. We explore different choices of $\mu_{\log f_{\rm Edd}}$ and $\sigma_{\log f_{\rm Edd}}$ to account for both asymmetric and more balanced accretion scenarios.

The resulting individual bolometric luminosities are converted into $K$-band magnitudes using the adopted bolometric correction and spectral assumptions described in Section~\ref{sec:comparison}. We then compute the corresponding flux ratio $F_2/F_1$ and determine, for each simulated binary and separation, the fraction of Monte Carlo realizations satisfying the \grav~ contrast criterion:
$
\frac{F_2}{F_1}\geq0.1.
$
The final \grav~ observable population is therefore determined jointly by the post-processed binary separation, the total $K$-band luminosity, and the assumed distribution of accretion luminosity between the two SMBHs.

The predicted binary AGN fraction is computed as
\begin{equation}
f_{\rm binary AGN} = 
\frac{N_{\rm binary AGN}}
{N_{\rm AGN}+N_{\rm binary AGN}},
\end{equation}
where $N_{\rm binary,AGN}$ is weighted by the probability of satisfying the \grav~ detectability criteria, including the flux-ratio detection probability, and $N_{\rm AGN}$ is the number of single AGN in the corresponding parent population.

This approach provides a consistent estimate of the \grav~-observable binary AGN fraction in the absence of spatially resolved information on the individual accretion states of the two SMBHs. It allows us to quantify how different assumptions about the relative accretion efficiency within the binary affect the predicted \grav~ detection rate while consistently accounting for the instrumental separation range, $K$-band sensitivity, and flux-ratio threshold.

\section{Comparison between observations and simulations}
\label{sec:comparison}
As discussed in Sect.~\ref{sec:binary_fraction}, the observed incidence of compact asymmetric structures is $1/20=5\%$, corresponding to $f_{\rm bin}<16.5\%$ at $1\sigma$ if NGC~3783 is considered a genuine binary, or $f_{\rm bin}<9.2\%$ if the observed asymmetry in NGC~3783 is instead attributed to non-binary circumnuclear emission.

\begin{figure}[b]
    \centering
    \includegraphics[width=1\linewidth]{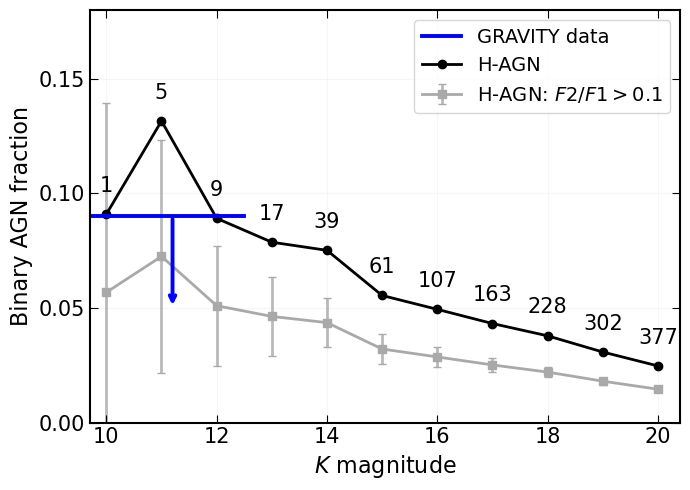}
    \caption{Binary AGN fraction for systems at $z<0.5$ with projected separations $<1$ kpc. Blue: upper limit derived from the GRAVITY observations. Black: prediction from \hagn. Grey: \hagn\ prediction after statistically assigning individual $f_{Edd}$ and requiring $F_2/F_1>0.1$.}
    \label{fig:fraction}
\end{figure}

To place this constraint in the context of galaxy evolution, we compare the observational upper limit with the binary AGN population extracted from the \hagn~simulation, applying a selection function matched to the \grav~sample in redshift, primary $K$-band magnitude, and projected separation. As discussed in Scialpi+26b (subm.), the overlap between cosmological simulations and the observational parameter space of dual and binary AGN depends strongly on simulation volume, spatial resolution, and selection criteria. In particular, the bright, low-redshift regime probed by the \grav~sample is challenging to reproduce. Among the simulations considered in Scialpi+26b (\textsc{L-Galaxies} \citep{Bonoli25_Lgalaxies}, \textsc{Astrid} \citep{Chen23b}, \textsc{Fable} \citep{Buttigieg25_fable}, \textsc{Horizon-AGN} \citep{Volonteri22}, and \textsc{Romulus} \citep{Saeedzadeh23_romulus_dual}), \hagn~is the only model producing systems within the relevant observational parameter space.
Even within \hagn~, however, the number of systems decreases rapidly toward the bright end, resulting in significant small-numbers for the $K<12.5$ selection considered here (Fig.~\ref{fig:fraction}). 
The comparison is further affected by the lack of individual SMBH accretion rates in Horizon-AGN, requiring the statistical reconstruction of the luminosity distribution between the two components described in Section~\ref{binary AGN}. The adopted treatment of dust obscuration also does not account for orientation-dependent effects, which may affect the predicted GRAVITY-detectable fraction.

We therefore apply this statistical accretion model to the simulated population and compare the resulting \grav~-observable fraction with the observational constraint, as shown in Fig.~\ref{fig:fraction}. The standard \hagn~prediction (black curve) lies above the observational upper limit (blue line) at $K<12.5$, whereas accounting for the individual accretion states and requiring $F_2/F_1>0.1$ (grey curve) reduces the predicted fraction, bringing it safely below the GRAVITY upper limit and into agreement within the uncertainties. Thus, when the individual accretion states and the observational flux-ratio selection are taken into account, the predicted and observed fractions are consistent within the assumptions of the model.



\section{Conclusion}
\label{sec:conclusion}

We presented a systematic search for compact secondary emission components in a sample of 20 nearby type 1 AGN observed with VLTI/\grav. By combining squared visibility amplitudes and closure phases, we investigated departures from centro-symmetric nuclear emission on milliarcsec scales, corresponding to separations of a few parsecs, depending on the source distance.

We developed a homogeneous analysis framework to identify candidate SMBH binaries while accounting for the main limitations of optical/infrared interferometric searches (Sect. \ref{sec:analysis}). Each AGN was first characterized through a single-component model describing the compact nuclear emission, followed by a companion search using binary models. Candidate detections were then assessed through injection--recovery simulations, which quantify the local sensitivity to secondary components, and through PSF-based diagnostics aimed at identifying interferometric artefacts.

Although several apparent secondary components are identified in the initial search, none remains consistent with a robust  detection after the complete validation procedure. Most candidates are explained by structured interferometric sidelobes, limited uv coverage, or low-level asymmetric nuclear emission as described in Sect.\ref{sec:no_binary_detection}. NGC~3783 represents the only source showing a persistent compact asymmetry, but its complex circumnuclear environment prevents an unambiguous interpretation, as asymmetric dust structures or outflow-related emission may produce similar interferometric signatures as demonstraded in Sect.\ref{sec:binary_candidates}.

We therefore identify one compact asymmetric component among the 20 targets, corresponding to an observed incidence of $5\%$. This result should not be interpreted as the intrinsic binary fraction, but rather as an empirical measure of the fraction of systems showing compact secondary-like emission within the spatial resolution and contrast limits of \grav.
If this candidate is interpreted as a genuine binary AGN detection, the Poissonian one-sided $1\sigma$ upper limit on the number of \grav-detectable binaries corresponds to $f_{\rm bin}<16.5\%$. On the other hand, if we interpret the observed asymmetry in NGC~3783 due to an AGN-heated gas cloud, we obtain a Poissonian one-sided $1\sigma$ upper limit of $f_{\rm bin}<9.2\%$ for \grav-detectable binaries among bright AGN at $z<0.5$ with $K<12.5$. 

We compare these observational results with the population of \grav-observable binary AGN predicted by the Horizon-AGN simulation, applying a selection function matched to the observed sample in redshift, $K$-band magnitude, projected separation, and flux ratio. The bright low-redshift regime probed by \grav is challenging to reproduce in cosmological simulations, and among the models considered in Scialpi+26b (subm.), Horizon-AGN is the only simulation producing systems within the relevant observational parameter space. Even within Horizon-AGN, the number of systems rapidly decreases toward the bright end, leading to significant small-number uncertainties for the $K<12.5$ selection. Moreover, the absence of individual SMBH accretion rates requires a statistical reconstruction of the luminosity distribution between the two components. Despite these limitations, the resulting \grav-observable binary fractions predicted by the simulations are broadly comparable to the observational constraints derived from the 20 AGN sample, providing consistency between the simulated population and the current interferometric results.

This work establishes a robust interferometric framework and provides the first GRAVITY-based upper limit on detectable parsec-scale binary AGN in bright local AGN. Infrared interferometry offers a unique avenue to probe the late stages of AGN pairing, bridging the gap between kpc-scale merger processes and the sub-parsec regime preceding GWs dominated evolution. Future observations with \grav+ and larger surveys will enable the study of fainter sources and statistically larger samples, improving the identification and characterization of binary AGN candidates. 
As a simple illustration of the sample size required by future interferometric surveys, if the true fraction is assumed to lie uniformly between zero and this upper limit, a sample of approximately 220 comparable AGN would be required to reach a $95\%$ probability of detecting at least one confirmed \grav-detectable binary AGN. This estimate is necessarily prior-dependent, but illustrates the need for substantially larger samples to turn the current upper limit into a statistically informative measurement of the compact binary AGN population.

Combined with complementary selection techniques, such as double-peaked emission-line searches and spectroscopic monitoring of radial velocity variability, this approach will provide a powerful framework for follow-up investigations of binary AGN candidates and for establishing the connection between galaxy mergers, SMBH binary formation, and future GW sources \citep[e.g.][]{Volonteri03,AmaroSeoane2023,Wang09_binary_doublepeak,Severgnini18_binary_doublepeak,Runnoe17,Runnoe25}.



\begin{acknowledgements}

This publication  was produced while attending the PhD program in PhD in Space Science and Technology at the University of Trento, Cycle XXXIX, with the support of a scholarship financed by the Ministerial Decree no. 118 of 2nd march 2023, based on the NRRP - funded by the European Union - NextGenerationEU - Mission 4 "Education and Research", Component 1 "Enhancement of the offer of educational services: from nurseries to universities” - Investment 4.1 “Extension of the number of research doctorates and innovative doctorates for public administration and cultural heritage”.   \\  

Based on observations collected at the European Southern Observatory under ESO Program IDs: 0109.B-0270 (PI: Davies) and 1103.B-0626 (PI: Sturm).

We thank the referee for the useful comments, which have helped us make the paper more complete.
We acknowledge financial support from the ASI-INAF agreement n. 2024-36-HH.1-2025. 
We thank the Galaxy and SMBH Formation, Growth, and Co-Evolution hiking Workshop in Nepal for the useful discussions on this paper.
LCH was supported by the China Manned Space Program (CMS-CSST-2025-A09) and the National Science Foundation of China (12233001). C.C., N.M. and P.G. acknowledge support through national funds by FCT – Fundação para a Ciência e a Tecnologia, I.P., Portugal, in the framework of the projects the Center for Astrophysics and Gravitation (CENTRA/IST/ULisboa) a through grants No. UID/PRR/00099/2025 (https://doi.org/10.54499/UID/PRR/00099/2025),  No. UID/00099/2025 (https://doi.org/10.54499/UID/00099/2025).
J.S-B. acknowledges the support received by the UNAM DGAPA-PAPIIT project IG-101025 and from the SECIHTI Ciencia de Frontera project CBF-2025-I-3033. R.G.M. acknowledges support from UNAM-PAPIIT project IN105225. 
The research leading to these results has received funding from the European Research Council (ERC) under the European Union's Horizon 2020 research and innovation program (project UniverScale, grant agreement 951549).
This work was supported by
- Paris Île-de-France Region;
- the French National Research Agency (ANR) under grant
ANR-23-EDIR-0003 (GRAFITY);
- the "Action Thématique Gravitation Références Astronomie Métrologie"
(ATGRAM), the "Action Thématique Phénomènes Extrêmes et Multimessagers"
(ATPEM), and the "Action Thématique Cosmologie et Galaxies" (ATCG), of
CNRS/INSU, with co-funding by CNRS/IN2P3, CNRS/INP, CEA and CNES.
\end{acknowledgements}

\bibliographystyle{aa}
\bibliography{references} 

\begin{thebibliography}{105}
\expandafter\ifx\csname natexlab\endcsname\relax\def\natexlab#1{#1}\fi

\bibitem[{{Agazie} {et~al.}(2023){Agazie}, {Anumarlapudi}, {Archibald}, {Arzoumanian}, {Baker}, {B{\'e}csy}, {Blecha}, {Brazier}, {Brook}, {Burke-Spolaor}, {Burnette}, {Case}, {Charisi}, {Chatterjee}, {Chatziioannou}, {Cheeseboro}, {Chen}, {Cohen}, {Cordes}, {Cornish}, {Crawford}, {Cromartie}, {Crowter}, {Cutler}, {Decesar}, {Degan}, {Demorest}, {Deng}, {Dolch}, {Drachler}, {Ellis}, {Ferrara}, {Fiore}, {Fonseca}, {Freedman}, {Garver-Daniels}, {Gentile}, {Gersbach}, {Glaser}, {Good}, {G{\"u}ltekin}, {Hazboun}, {Hourihane}, {Islo}, {Jennings}, {Johnson}, {Jones}, {Kaiser}, {Kaplan}, {Kelley}, {Kerr}, {Key}, {Klein}, {Laal}, {Lam}, {Lamb}, {Lazio}, {Lewandowska}, {Littenberg}, {Liu}, {Lommen}, {Lorimer}, {Luo}, {Lynch}, {Ma}, {Madison}, {Mattson}, {McEwen}, {McKee}, {McLaughlin}, {McMann}, {Meyers}, {Meyers}, {Mingarelli}, {Mitridate}, {Natarajan}, {Ng}, {Nice}, {Ocker}, {Olum}, {Pennucci}, {Perera}, {Petrov}, {Pol}, {Radovan}, {Ransom}, {Ray}, {Romano}, {Sardesai}, {Schmiedekamp}, {Schmiedekamp}, {Schmitz},
  {Schult}, {Shapiro-Albert}, {Siemens}, {Simon}, {Siwek}, {Stairs}, {Stinebring}, {Stovall}, {Sun}, {Susobhanan}, {Swiggum}, {Taylor}, {Taylor}, {Turner}, {Unal}, {Vallisneri}, {van Haasteren}, {Vigeland}, {Wahl}, {Wang}, {Witt}, {Young}, \& {Nanograv Collaboration}}]{Agazie2023a}
{Agazie}, G., {Anumarlapudi}, A., {Archibald}, A.~M., {et~al.} 2023, \apjl, 951, L8

\bibitem[{{Amaro-Seoane} {et~al.}(2023){Amaro-Seoane}, {Andrews}, {Arca Sedda}, {Askar}, {Baghi}, {Balasov}, {Bartos}, {Bavera}, {Bellovary}, {Berry}, {Berti}, {Bianchi}, {Blecha}, {Blondin}, {Bogdanovi{\'c}}, {Boissier}, {Bonetti}, {Bonoli}, {Bortolas}, {Breivik}, {Capelo}, {Caramete}, {Cattorini}, {Charisi}, {Chaty}, {Chen}, {Chru{\'s}li{\'n}ska}, {Chua}, {Church}, {Colpi}, {D'Orazio}, {Danielski}, {Davies}, {Dayal}, {De Rosa}, {Derdzinski}, {Destounis}, {Dotti}, {Du{\c{t}}an}, {Dvorkin}, {Fabj}, {Foglizzo}, {Ford}, {Fouvry}, {Franchini}, {Fragos}, {Fryer}, {Gaspari}, {Gerosa}, {Graziani}, {Groot}, {Habouzit}, {Haggard}, {Haiman}, {Han}, {Istrate}, {Johansson}, {Khan}, {Kimpson}, {Kokkotas}, {Kong}, {Korol}, {Kremer}, {Kupfer}, {Lamberts}, {Larson}, {Lau}, {Liu}, {Lloyd-Ronning}, {Lodato}, {Lupi}, {Ma}, {Maccarone}, {Mandel}, {Mangiagli}, {Mapelli}, {Mathis}, {Mayer}, {McGee}, {McKernan}, {Miller}, {Mota}, {Mumpower}, {Nasim}, {Nelemans}, {Noble}, {Pacucci}, {Panessa}, {Paschalidis}, {Pfister}, {Porquet},
  {Quenby}, {Ricarte}, {R{\"o}pke}, {Regan}, {Rosswog}, {Ruiter}, {Ruiz}, {Runnoe}, {Schneider}, {Schnittman}, {Secunda}, {Sesana}, {Seto}, {Shao}, {Shapiro}, {Sopuerta}, {Stone}, {Suvorov}, {Tamanini}, {Tamfal}, {Tauris}, {Temmink}, {Tomsick}, {Toonen}, {Torres-Orjuela}, {Toscani}, {Tsokaros}, {Unal}, {V{\'a}zquez-Aceves}, {Valiante}, {van Putten}, {van Roestel}, {Vignali}, {Volonteri}, {Wu}, {Younsi}, {Yu}, {Zane}, {Zwick}, {Antonini}, {Baibhav}, {Barausse}, {Bonilla Rivera}, {Branchesi}, {Branduardi-Raymont}, {Burdge}, {Chakraborty}, {Cuadra}, {Dage}, {Davis}, {de Mink}, {Decarli}, {Doneva}, {Escoffier}, {Gandhi}, {Haardt}, {Lousto}, {Nissanke}, {Nordhaus}, {O'Shaughnessy}, {Portegies Zwart}, {Pound}, {Schussler}, {Sergijenko}, {Spallicci}, {Vernieri}, \& {Vigna-G{\'o}mez}}]{AmaroSeoane2023}
{Amaro-Seoane}, P., {Andrews}, J., {Arca Sedda}, M., {et~al.} 2023, Living Reviews in Relativity, 26, 2

\bibitem[{{Aubert} {et~al.}(2004){Aubert}, {Pichon}, \& {Colombi}}]{aubertetal04}
{Aubert}, D., {Pichon}, C., \& {Colombi}, S. 2004, \mnras, 352, 376

\bibitem[{{Battistini} {et~al.}(2026){Battistini}, {De Rosa}, {Severgnini}, {Vignali}, {Singh}, {Capelo}, {Bertola}, {Bianchi}, {D'Amato}, {Guainazzi}, {La Franca}, {Lamperti}, {Mannucci}, {Parvatikar}, {Piconcelli}, {Ricci}, {Rigamonti}, {Scialpi}, \& {Vittoria Zanchettin}}]{Battistini26}
{Battistini}, L., {De Rosa}, A., {Severgnini}, P., {et~al.} 2026, arXiv e-prints, arXiv:2604.14900

\bibitem[{{Becker} {et~al.}(1995){Becker}, {White}, \& {Helfand}}]{Becker95}
{Becker}, R.~H., {White}, R.~L., \& {Helfand}, D.~J. 1995, \apj, 450, 559

\bibitem[{{Begelman} {et~al.}(1980){Begelman}, {Blandford}, \& {Rees}}]{Begelman80}
{Begelman}, M.~C., {Blandford}, R.~D., \& {Rees}, M.~J. 1980, \nat, 287, 307

\bibitem[{{Bogdanovic} {et~al.}(2021){Bogdanovic}, {Nguyen}, {Runnoe}, {Taylor}, {Sesana}, {Eracleous}, \& {Sigurdsson}}]{Bogdanovic21}
{Bogdanovic}, T., {Nguyen}, K., {Runnoe}, J.~C., {et~al.} 2021, in American Astronomical Society Meeting Abstracts, Vol. 237, American Astronomical Society Meeting Abstracts \#237, 518.05

\bibitem[{{Bonoli} {et~al.}(2025){Bonoli}, {Izquierdo-Villalba}, {Spinoso}, {Colpi}, {Sesana}, {Polkas}, \& {Springel}}]{Bonoli25_Lgalaxies}
{Bonoli}, S., {Izquierdo-Villalba}, D., {Spinoso}, D., {et~al.} 2025, arXiv e-prints, arXiv:2509.12325

\bibitem[{{Burke-Spolaor} {et~al.}(2019){Burke-Spolaor}, {Taylor}, {Charisi}, {Dolch}, {Hazboun}, {Holgado}, {Kelley}, {Lazio}, {Madison}, {McMann}, {Mingarelli}, {Rasskazov}, {Siemens}, {Simon}, \& {Smith}}]{Burke-Spolaor19}
{Burke-Spolaor}, S., {Taylor}, S.~R., {Charisi}, M., {et~al.} 2019, \aapr, 27, 5

\bibitem[{{Burtscher} {et~al.}(2013){Burtscher}, {Meisenheimer}, {Tristram}, {Jaffe}, {H{\"o}nig}, {Davies}, {Kishimoto}, {Pott}, {R{\"o}ttgering}, {Schartmann}, {Weigelt}, \& {Wolf}}]{Burtscher13}
{Burtscher}, L., {Meisenheimer}, K., {Tristram}, K.~R.~W., {et~al.} 2013, \aap, 558, A149

\bibitem[{{Buttigieg} {et~al.}(2025){Buttigieg}, {Sijacki}, {Moore}, \& {Bourne}}]{Buttigieg25_fable}
{Buttigieg}, S., {Sijacki}, D., {Moore}, C.~J., \& {Bourne}, M.~A. 2025, \mnras, 542, 2019

\bibitem[{{Cardaci} {et~al.}(2009){Cardaci}, {Santos-Lle{\'o}}, {Krongold}, {H{\"a}gele}, {D{\'\i}az}, \& {Rodr{\'\i}guez-Pascual}}]{Cardaci09}
{Cardaci}, M.~V., {Santos-Lle{\'o}}, M., {Krongold}, Y., {et~al.} 2009, \aap, 505, 541

\bibitem[{{Charisi}(2022)}]{charisi22}
{Charisi}, M. 2022, in APS Meeting Abstracts, Vol. 2022, APS April Meeting Abstracts, X14.005

\bibitem[{{Charisi} {et~al.}(2016){Charisi}, {Bartos}, {Haiman}, {Price-Whelan}, {Graham}, {Bellm}, {Laher}, \& {M{\'a}rka}}]{Charisi16}
{Charisi}, M., {Bartos}, I., {Haiman}, Z., {et~al.} 2016, \mnras, 463, 2145

\bibitem[{{Charisi} {et~al.}(2018){Charisi}, {Haiman}, {Schiminovich}, \& {D'Orazio}}]{Charisi18}
{Charisi}, M., {Haiman}, Z., {Schiminovich}, D., \& {D'Orazio}, D.~J. 2018, \mnras, 476, 4617

\bibitem[{{Chen} {et~al.}(2023){Chen}, {Di Matteo}, {Ni}, {Tremmel}, {DeGraf}, {Shen}, {Holgado}, {Bird}, {Croft}, \& {Feng}}]{Chen23b}
{Chen}, N., {Di Matteo}, T., {Ni}, Y., {et~al.} 2023, \mnras, 522, 1895

\bibitem[{{Chen} {et~al.}(2022{\natexlab{a}}){Chen}, {Ni}, {Tremmel}, {Di Matteo}, {Bird}, {DeGraf}, \& {Feng}}]{2022MNRAS.510..531C}
{Chen}, N., {Ni}, Y., {Tremmel}, M., {et~al.} 2022{\natexlab{a}}, \mnras, 510, 531

\bibitem[{{Chen} {et~al.}(2022{\natexlab{b}}){Chen}, {Hwang}, {Shen}, {Liu}, {Zakamska}, {Yang}, \& {Li}}]{Chen22a}
{Chen}, Y.-C., {Hwang}, H.-C., {Shen}, Y., {et~al.} 2022{\natexlab{b}}, \apj, 925, 162

\bibitem[{{Colpi} {et~al.}(2024){Colpi}, {Danzmann}, {Hewitson}, {Holley-Bockelmann}, {Jetzer}, {Nelemans}, {Petiteau}, {Shoemaker}, {Sopuerta}, {Stebbins}, {Tanvir}, {Ward}, {Weber}, {Thorpe}, {Daurskikh}, {Deep}, {Fern{\'a}ndez N{\'u}{\~n}ez}, {Garc{\'\i}a Marirrodriga}, {Gehler}, {Halain}, {Jennrich}, {Lammers}, {Larra{\~n}aga}, {Lieser}, {L{\"u}tzgendorf}, {Martens}, {Mondin}, {Piris Ni{\~n}o}, {Amaro-Seoane}, {Arca Sedda}, {Auclair}, {Babak}, {Baghi}, {Baibhav}, {Baker}, {Bayle}, {Berry}, {Berti}, {Boileau}, {Bonetti}, {Brito}, {Buscicchio}, {Calcagni}, {Capelo}, {Caprini}, {Caputo}, {Castelli}, {Chen}, {Chen}, {Chua}, {Davies}, {Derdzinski}, {Domcke}, {Doneva}, {Dvorkin}, {Mar{\'\i}a Ezquiaga}, {Gair}, {Haiman}, {Harry}, {Hartwig}, {Hees}, {Heffernan}, {Husa}, {Izquierdo-Villalba}, {Karnesis}, {Klein}, {Korol}, {Korsakova}, {Kupfer}, {Laghi}, {Lamberts}, {Larson}, {Le Jeune}, {Lewicki}, {Littenberg}, {Madge}, {Mangiagli}, {Marsat}, {Vilchez}, {Maselli}, {Mathews}, {van de Meent}, {Muratore}, {Nardini},
  {Pani}, {Peloso}, {Pieroni}, {Pound}, {Quelquejay-Leclere}, {Ricciardone}, {Rossi}, {Sartirana}, {Savalle}, {Sberna}, {Sesana}, {Shoemaker}, {Slutsky}, {Sotiriou}, {Speri}, {Staab}, {Steer}, {Tamanini}, {Tasinato}, {Torrado}, {Torres-Orjuela}, {Toubiana}, {Vallisneri}, {Vecchio}, {Volonteri}, {Yagi}, \& {Zwick}}]{Colpi2024arxiv}
{Colpi}, M., {Danzmann}, K., {Hewitson}, M., {et~al.} 2024, arXiv e-prints, arXiv:2402.07571

\bibitem[{{Condon} {et~al.}(1998){Condon}, {Cotton}, {Greisen}, {Yin}, {Perley}, {Taylor}, \& {Broderick}}]{Condon98}
{Condon}, J.~J., {Cotton}, W.~D., {Greisen}, E.~W., {et~al.} 1998, \aj, 115, 1693

\bibitem[{De~Rosa {et~al.}(2020)De~Rosa, Vignali, Bogdanović, Capelo, Charisi, Dotti, Husemann, Lusso, Mayer, Paragi, Runnoe, Sesana, Steinborn, Bianchi, Colpi, Del~Valle, Frey, Gabányi, Giustini, Guainazzi, Haiman, Herrera~Ruiz, Herrero-Illana, Iwasawa, Komossa, Lena, Loiseau, Perez-Torres, Piconcelli, \& Volonteri}]{derosa20}
De~Rosa, A., Vignali, C., Bogdanović, T., {et~al.} 2020, arXiv e-prints, arXiv:2001.06293, \_eprint: 2001.06293

\bibitem[{{De Rosa} {et~al.}(2023){De Rosa}, {Vignali}, {Severgnini}, {Bianchi}, {Bogdanovi{\'c}}, {Charisi}, {Guainazzi}, {Haiman}, {Komossa}, {Paragi}, {Perez-Torres}, {Piconcelli}, {Ducci}, {Parvatikar}, \& {Serafinelli}}]{DeRosa2023}
{De Rosa}, A., {Vignali}, C., {Severgnini}, P., {et~al.} 2023, \mnras, 519, 5149

\bibitem[{{DESI Collaboration} {et~al.}(2025){DESI Collaboration}, {Abdul-Karim}, {Adame}, {Aguado}, {Aguilar}, {Ahlen}, {Alam}, {Aldering}, {Alexander}, {Alfarsy}, {Allen}, {Allende Prieto}, {Alves}, {Anand}, {Andrade}, {Armengaud}, {Avila}, {Aviles}, {Awan}, {Bailey}, {Baleato Lizancos}, {Ballester}, {Bault}, {Bautista}, {BenZvi}, {Beraldo e Silva}, {Bermejo-Climent}, {Beutler}, {Bianchi}, {Blake}, {Blum}, {Bolton}, {Bonici}, {Brieden}, {Brodzeller}, {Brooks}, {Buckley-Geer}, {Burtin}, {Canning}, {Carnero Rosell}, {Carr}, {Carrilho}, {Casas}, {Castander}, {Cereskaite}, {Cervantes-Cota}, {Chaussidon}, {Chaves-Montero}, {Chen}, {Chen}, {Claybaugh}, {Cole}, {Cooper}, {Cousinou}, {Cuceu}, {Davis}, {Dawson}, {de Belsunce}, {de la Cruz}, {de la Macorra}, {de Mattia}, {Deiosso}, {Della Costa}, {Demina}, {Demirbozan}, {DeRose}, {Dey}, {Dey}, {Ding}, {Ding}, {Doel}, {Douglass}, {Dowicz}, {Ebina}, {Edelstein}, {Eisenstein}, {Elbers}, {Emas}, {Escoffier}, {Fagrelius}, {Fan}, {Fanning}, {Fawcett},
  {Fern\textbackslash'andez-Garc\textbackslash'ia}, {Ferraro}, {Findlay}, {Font-Ribera}, {Forero-Romero}, {Forero-S\textbackslash'anchez}, {Frenk}, {G\textbackslash''ansicke}, {Galbany}, {Garc\textbackslash'ia-Bellido}, {Garcia-Quintero}, \& {Garrison}}]{DESI25_arxive}
{DESI Collaboration}, {Abdul-Karim}, M., {Adame}, A.~G., {et~al.} 2025, arXiv e-prints, arXiv:2503.14745

\bibitem[{{Di Matteo} {et~al.}(2005){Di Matteo}, {Springel}, \& {Hernquist}}]{DiMatteo05}
{Di Matteo}, T., {Springel}, V., \& {Hernquist}, L. 2005, in Growing Black Holes: Accretion in a Cosmological Context, ed. A.~{Merloni}, S.~{Nayakshin}, \& R.~A. {Sunyaev}, 340--345

\bibitem[{{D'Orazio} \& {Charisi}(2023)}]{DOrazioCharisi23}
{D'Orazio}, D.~J. \& {Charisi}, M. 2023, arXiv e-prints, arXiv:2310.16896

\bibitem[{{Dubois} {et~al.}(2013){Dubois}, {Pichon}, {Devriendt}, {Silk}, {Haehnelt}, {Kimm}, \& {Slyz}}]{Dubois2013}
{Dubois}, Y., {Pichon}, C., {Devriendt}, J., {et~al.} 2013, \mnras, 428, 2885

\bibitem[{{Dubois} {et~al.}(2014){Dubois}, {Pichon}, {Welker}, {Le Borgne}, {Devriendt}, {Laigle}, {Codis}, {Pogosyan}, {Arnouts}, {Benabed}, {Bertin}, {Blaizot}, {Bouchet}, {Cardoso}, {Colombi}, {de Lapparent}, {Desjacques}, {Gavazzi}, {Kassin}, {Kimm}, {McCracken}, {Milliard}, {Peirani}, {Prunet}, {Rouberol}, {Silk}, {Slyz}, {Sousbie}, {Teyssier}, {Tresse}, {Treyer}, {Vibert}, \& {Volonteri}}]{Dubois14}
{Dubois}, Y., {Pichon}, C., {Welker}, C., {et~al.} 2014, \mnras, 444, 1453

\bibitem[{{Eisenhauer} {et~al.}(2023){Eisenhauer}, {Monnier}, \& {Pfuhl}}]{Eisenhauer23}
{Eisenhauer}, F., {Monnier}, J.~D., \& {Pfuhl}, O. 2023, \araa, 61, 237

\bibitem[{{EPTA Collaboration} {et~al.}(2023){EPTA Collaboration}, {InPTA Collaboration}, {Antoniadis}, {Arumugam}, {Arumugam}, {Babak}, {Bagchi}, {Bak Nielsen}, {Bassa}, {Bathula}, {Berthereau}, {Bonetti}, {Bortolas}, {Brook}, {Burgay}, {Caballero}, {Chalumeau}, {Champion}, {Chanlaridis}, {Chen}, {Cognard}, {Dandapat}, {Deb}, {Desai}, {Desvignes}, {Dhanda-Batra}, {Dwivedi}, {Falxa}, {Ferdman}, {Franchini}, {Gair}, {Goncharov}, {Gopakumar}, {Graikou}, {Grie{\ss}meier}, {Guillemot}, {Guo}, {Gupta}, {Hisano}, {Hu}, {Iraci}, {Izquierdo-Villalba}, {Jang}, {Jawor}, {Janssen}, {Jessner}, {Joshi}, {Kareem}, {Karuppusamy}, {Keane}, {Keith}, {Kharbanda}, {Kikunaga}, {Kolhe}, {Kramer}, {Krishnakumar}, {Lackeos}, {Lee}, {Liu}, {Liu}, {Lyne}, {McKee}, {Maan}, {Main}, {Mickaliger}, {Ni{\c{t}}u}, {Nobleson}, {Paladi}, {Parthasarathy}, {Perera}, {Perrodin}, {Petiteau}, {Porayko}, {Possenti}, {Prabu}, {Quelquejay Leclere}, {Rana}, {Samajdar}, {Sanidas}, {Sesana}, {Shaifullah}, {Singha}, {Speri}, {Spiewak}, {Srivastava},
  {Stappers}, {Surnis}, {Susarla}, {Susobhanan}, {Takahashi}, {Tarafdar}, {Theureau}, {Tiburzi}, {van der Wateren}, {Vecchio}, {Venkatraman Krishnan}, {Verbiest}, {Wang}, {Wang}, \& {Wu}}]{EPTACollaboration2023}
{EPTA Collaboration}, {InPTA Collaboration}, {Antoniadis}, J., {et~al.} 2023, \aap, 678, A50

\bibitem[{{Euclid Collaboration: Mellier} {et~al.}(2024){Euclid Collaboration: Mellier}, {Abdurro'uf}, {Acevedo~Barroso}, {et~al.}}]{Euclid_Overview}
{Euclid Collaboration: Mellier}, Y., {Abdurro'uf}, {Acevedo~Barroso}, J., {et~al.} 2024, \aap, accepted, arXiv:2405.13491

\bibitem[{{Fu} {et~al.}(2015{\natexlab{a}}){Fu}, {Myers}, {Djorgovski}, {Yan}, {Wrobel}, \& {Stockton}}]{Fu15_a}
{Fu}, H., {Myers}, A.~D., {Djorgovski}, S.~G., {et~al.} 2015{\natexlab{a}}, \apj, 799, 72

\bibitem[{{Fu} {et~al.}(2015{\natexlab{b}}){Fu}, {Wrobel}, {Myers}, {Djorgovski}, \& {Yan}}]{Fu15_b}
{Fu}, H., {Wrobel}, J.~M., {Myers}, A.~D., {Djorgovski}, S.~G., \& {Yan}, L. 2015{\natexlab{b}}, \apjl, 815, L6

\bibitem[{{Gaia Collaboration} {et~al.}(2021){Gaia Collaboration}, {Brown}, {Vallenari}, {Prusti}, {de Bruijne}, {Babusiaux}, {Biermann}, {Creevey}, {Evans}, {Eyer}, {Hutton}, {Jansen}, {Jordi}, {Klioner}, {Lammers}, {Lindegren}, {Luri}, {Mignard}, {Panem}, {Pourbaix}, {Randich}, {Sartoretti}, {Soubiran}, {Walton}, {Arenou}, {Bailer-Jones}, {Bastian}, {Cropper}, {Drimmel}, {Katz}, {Lattanzi}, {van Leeuwen}, {Bakker}, {Cacciari}, {Casta{\~n}eda}, {De Angeli}, {Ducourant}, {Fabricius}, {Fouesneau}, {Fr{\'e}mat}, {Guerra}, {Guerrier}, {Guiraud}, {Jean-Antoine Piccolo}, {Masana}, {Messineo}, {Mowlavi}, {Nicolas}, {Nienartowicz}, {Pailler}, {Panuzzo}, {Riclet}, {Roux}, {Seabroke}, {Sordo}, {Tanga}, {Th{\'e}venin}, {Gracia-Abril}, {Portell}, {Teyssier}, {Altmann}, {Andrae}, {Bellas-Velidis}, {Benson}, {Berthier}, {Blomme}, {Brugaletta}, {Burgess}, {Busso}, {Carry}, {Cellino}, {Cheek}, {Clementini}, {Damerdji}, {Davidson}, {Delchambre}, {Dell'Oro}, {Fern{\'a}ndez-Hern{\'a}ndez}, {Galluccio}, {Garc{\'\i}a-Lario},
  {Garcia-Reinaldos}, {Gonz{\'a}lez-N{\'u}{\~n}ez}, {Gosset}, {Haigron}, {Halbwachs}, {Hambly}, {Harrison}, {Hatzidimitriou}, {Heiter}, {Hern{\'a}ndez}, {Hestroffer}, {Hodgkin}, {Holl}, {Jan{\ss}en}, {Jevardat de Fombelle}, {Jordan}, {Krone-Martins}, {Lanzafame}, {L{\"o}ffler}, {Lorca}, {Manteiga}, {Marchal}, {Marrese}, {Moitinho}, {Mora}, {Muinonen}, {Osborne}, {Pancino}, {Pauwels}, {Petit}, {Recio-Blanco}, {Richards}, {Riello}, {Rimoldini}, {Robin}, {Roegiers}, {Rybizki}, {Sarro}, {Siopis}, {Smith}, {Sozzetti}, {Ulla}, {Utrilla}, {van Leeuwen}, {van Reeven}, {Abbas}, {Abreu Aramburu}, {Accart}, {Aerts}, {Aguado}, {Ajaj}, {Altavilla}, {{\'A}lvarez}, {{\'A}lvarez Cid-Fuentes}, {Alves}, {Anderson}, {Anglada Varela}, {Antoja}, {Audard}, {Baines}, {Baker}, {Balaguer-N{\'u}{\~n}ez}, {Balbinot}, {Balog}, {Barache}, {Barbato}, {Barros}, {Barstow}, {Bartolom{\'e}}, {Bassilana}, {Bauchet}, {Baudesson-Stella}, {Becciani}, {Bellazzini}, {Bernet}, {Bertone}, {Bianchi}, {Blanco-Cuaresma}, {Boch}, {Bombrun}, {Bossini},
  {Bouquillon}, {Bragaglia}, {Bramante}, {Breedt}, {Bressan}, {Brouillet}, {Bucciarelli}, {Burlacu}, {Busonero}, {Butkevich}, {Buzzi}, {Caffau}, {Cancelliere}, {C{\'a}novas}, {Cantat-Gaudin}, {Carballo}, {Carlucci}, {Carnerero}, {Carrasco}, {Casamiquela}, {Castellani}, {Castro-Ginard}, {Castro Sampol}, {Chaoul}, {Charlot}, {Chemin}, {Chiavassa}, {Cioni}, {Comoretto}, {Cooper}, {Cornez}, {Cowell}, {Crifo}, {Crosta}, {Crowley}, {Dafonte}, {Dapergolas}, {David}, {David}, {de Laverny}, {De Luise}, {De March}, {De Ridder}, {de Souza}, {de Teodoro}, {de Torres}, {del Peloso}, {del Pozo}, {Delbo}, {Delgado}, {Delgado}, {Delisle}, {Di Matteo}, {Diakite}, {Diener}, {Distefano}, {Dolding}, {Eappachen}, {Edvardsson}, {Enke}, {Esquej}, {Fabre}, {Fabrizio}, {Faigler}, {Fedorets}, {Fernique}, {Fienga}, {Figueras}, {Fouron}, {Fragkoudi}, {Fraile}, {Franke}, {Gai}, {Garabato}, {Garcia-Gutierrez}, {Garc{\'\i}a-Torres}, {Garofalo}, {Gavras}, {Gerlach}, {Geyer}, {Giacobbe}, {Gilmore}, {Girona}, {Giuffrida}, {Gomel}, {Gomez},
  {Gonzalez-Santamaria}, {Gonz{\'a}lez-Vidal}, {Granvik}, {Guti{\'e}rrez-S{\'a}nchez}, {Guy}, {Hauser}, {Haywood}, {Helmi}, {Hidalgo}, {Hilger}, {H{\l}adczuk}, {Hobbs}, {Holland}, {Huckle}, {Jasniewicz}, {Jonker}, {Juaristi Campillo}, {Julbe}, {Karbevska}, {Kervella}, {Khanna}, {Kochoska}, {Kontizas}, {Kordopatis}, {Korn}, {Kostrzewa-Rutkowska}, {Kruszy{\'n}ska}, {Lambert}, {Lanza}, {Lasne}, {Le Campion}, {Le Fustec}, {Lebreton}, {Lebzelter}, {Leccia}, {Leclerc}, {Lecoeur-Taibi}, {Liao}, {Licata}, {Lindstr{\o}m}, {Lister}, {Livanou}, {Lobel}, {Madrero Pardo}, {Managau}, {Mann}, {Marchant}, {Marconi}, {Marcos Santos}, {Marinoni}, {Marocco}, {Marshall}, {Martin Polo}, {Mart{\'\i}n-Fleitas}, {Masip}, {Massari}, {Mastrobuono-Battisti}, {Mazeh}, {McMillan}, {Messina}, {Michalik}, {Millar}, {Mints}, {Molina}, {Molinaro}, {Moln{\'a}r}, {Montegriffo}, {Mor}, {Morbidelli}, {Morel}, {Morris}, {Mulone}, {Munoz}, {Muraveva}, {Murphy}, {Musella}, {Noval}, {Ord{\'e}novic}, {Orr{\`u}}, {Osinde}, {Pagani}, {Pagano},
  {Palaversa}, {Palicio}, {Panahi}, {Pawlak}, {Pe{\~n}alosa Esteller}, {Penttil{\"a}}, {Piersimoni}, {Pineau}, {Plachy}, {Plum}, {Poggio}, {Poretti}, {Poujoulet}, {Pr{\v{s}}a}, {Pulone}, {Racero}, {Ragaini}, {Rainer}, {Raiteri}, {Rambaux}, {Ramos}, {Ramos-Lerate}, {Re Fiorentin}, {Regibo}, {Reyl{\'e}}, {Ripepi}, {Riva}, {Rixon}, {Robichon}, {Robin}, {Roelens}, {Rohrbasser}, {Romero-G{\'o}mez}, {Rowell}, {Royer}, {Rybicki}, {Sadowski}, {Sagrist{\`a} Sell{\'e}s}, {Sahlmann}, {Salgado}, {Salguero}, {Samaras}, {Sanchez Gimenez}, {Sanna}, {Santove{\~n}a}, {Sarasso}, {Schultheis}, {Sciacca}, {Segol}, {Segovia}, {S{\'e}gransan}, {Semeux}, {Shahaf}, {Siddiqui}, {Siebert}, {Siltala}, {Slezak}, {Smart}, {Solano}, {Solitro}, {Souami}, {Souchay}, {Spagna}, {Spoto}, {Steele}, {Steidelm{\"u}ller}, {Stephenson}, {S{\"u}veges}, {Szabados}, {Szegedi-Elek}, {Taris}, {Tauran}, {Taylor}, {Teixeira}, {Thuillot}, {Tonello}, {Torra}, {Torra}, {Turon}, {Unger}, {Vaillant}, {van Dillen}, {Vanel}, {Vecchiato}, {Viala}, {Vicente},
  {Voutsinas}, {Weiler}, {Wevers}, {Wyrzykowski}, {Yoldas}, {Yvard}, {Zhao}, {Zorec}, {Zucker}, {Zurbach}, \& {Zwitter}}]{Gaia_EDR3_1}
{Gaia Collaboration}, {Brown}, A.~G.~A., {Vallenari}, A., {et~al.} 2021, \aap, 649, A1

\bibitem[{{Glikman} {et~al.}(2023){Glikman}, {Langgin}, {Johnstone}, {Yoon}, {Comerford}, {Simmons}, {Stacey}, {Lacy}, \& {O'Meara}}]{Glikman23}
{Glikman}, E., {Langgin}, R., {Johnstone}, M.~A., {et~al.} 2023, \apjl, 951, L18

\bibitem[{{GRAVITY Collaboration} {et~al.}(2017){GRAVITY Collaboration}, {Abuter}, {Accardo}, {Amorim}, {Anugu}, {{\'A}vila}, {Azouaoui}, {Benisty}, {Berger}, {Blind}, {Bonnet}, {Bourget}, {Brandner}, {Brast}, {Buron}, {Burtscher}, {Cassaing}, {Chapron}, {Choquet}, {Cl{\'e}net}, {Collin}, {Coud{\'e} Du Foresto}, {de Wit}, {de Zeeuw}, {Deen}, {Delplancke-Str{\"o}bele}, {Dembet}, {Derie}, {Dexter}, {Duvert}, {Ebert}, {Eckart}, {Eisenhauer}, {Esselborn}, {F{\'e}dou}, {Finger}, {Garcia}, {Garcia Dabo}, {Garcia Lopez}, {Gendron}, {Genzel}, {Gillessen}, {Gonte}, {Gordo}, {Grould}, {Gr{\"o}zinger}, {Guieu}, {Haguenauer}, {Hans}, {Haubois}, {Haug}, {Haussmann}, {Henning}, {Hippler}, {Horrobin}, {Huber}, {Hubert}, {Hubin}, {Hummel}, {Jakob}, {Janssen}, {Jochum}, {Jocou}, {Kaufer}, {Kellner}, {Kendrew}, {Kern}, {Kervella}, {Kiekebusch}, {Klein}, {Kok}, {Kolb}, {Kulas}, {Lacour}, {Lapeyr{\`e}re}, {Lazareff}, {Le Bouquin}, {L{\`e}na}, {Lenzen}, {L{\'e}v{\^e}que}, {Lippa}, {Magnard}, {Mehrgan}, {Mellein}, {M{\'e}rand},
  {Moreno-Ventas}, {Moulin}, {M{\"u}ller}, {M{\"u}ller}, {Neumann}, {Oberti}, {Ott}, {Pallanca}, {Panduro}, {Pasquini}, {Paumard}, {Percheron}, {Perraut}, {Perrin}, {Pfl{\"u}ger}, {Pfuhl}, {Phan Duc}, {Plewa}, {Popovic}, {Rabien}, {Ram{\'\i}rez}, {Ramos}, {Rau}, {Riquelme}, {Rohloff}, {Rousset}, {Sanchez-Bermudez}, {Scheithauer}, {Sch{\"o}ller}, {Schuhler}, {Spyromilio}, {Straubmeier}, {Sturm}, {Suarez}, {Tristram}, {Ventura}, {Vincent}, {Waisberg}, {Wank}, {Weber}, {Wieprecht}, {Wiest}, {Wiezorrek}, {Wittkowski}, {Woillez}, {Wolff}, {Yazici}, {Ziegler}, \& {Zins}}]{Gravity+17_1}
{GRAVITY Collaboration}, {Abuter}, R., {Accardo}, M., {et~al.} 2017, \aap, 602, A94

\bibitem[{{GRAVITY Collaboration} {et~al.}(2021){GRAVITY Collaboration}, {Amorim}, {Baub{\"o}ck}, {Brandner}, {Bolzer}, {Cl{\'e}net}, {Davies}, {de Zeeuw}, {Dexter}, {Drescher}, {Eckart}, {Eisenhauer}, {F{\"o}rster Schreiber}, {Gao}, {Garcia}, {Genzel}, {Gillessen}, {Gratadour}, {H{\"o}nig}, {Kaltenbrunner}, {Kishimoto}, {Lacour}, {Lutz}, {Millour}, {Netzer}, {Ott}, {Paumard}, {Perraut}, {Perrin}, {Peterson}, {Petrucci}, {Pfuhl}, {Prieto}, {Rouan}, {Sanchez-Bermudez}, {Shangguan}, {Shimizu}, {Schartmann}, {Stadler}, {Sternberg}, {Straub}, {Straubmeier}, {Sturm}, {Tacconi}, {Tristram}, {Vermot}, {von Fellenberg}, {Waisberg}, {Widmann}, \& {Woillez}}]{gravity_taro_NGC3783}
{GRAVITY Collaboration}, {Amorim}, A., {Baub{\"o}ck}, M., {et~al.} 2021, \aap, 648, A117

\bibitem[{{GRAVITY Collaboration} {et~al.}(2024){GRAVITY Collaboration}, {Amorim}, {Bourdarot}, {Brandner}, {Cao}, {Cl{\'e}net}, {Davies}, {de Zeeuw}, {Dexter}, {Drescher}, {Eckart}, {Eisenhauer}, {Fabricius}, {Feuchtgruber}, {F{\"o}rster Schreiber}, {Garcia}, {Genzel}, {Gillessen}, {Gratadour}, {H{\"o}nig}, {Kishimoto}, {Lacour}, {Lutz}, {Millour}, {Netzer}, {Ott}, {Perraut}, {Perrin}, {Peterson}, {Petrucci}, {Pfuhl}, {Prieto}, {Rabien}, {Rouan}, {Santos}, {Shangguan}, {Shimizu}, {Sternberg}, {Straubmeier}, {Sturm}, {Tacconi}, {Tristram}, {Widmann}, \& {Woillez}}]{Cao24}
{GRAVITY Collaboration}, {Amorim}, A., {Bourdarot}, G., {et~al.} 2024, \aap, 690, A76

\bibitem[{{GRAVITY Collaboration} {et~al.}(2020{\natexlab{a}}){GRAVITY Collaboration}, {Dexter}, {Shangguan}, {H{\"o}nig}, {Kishimoto}, {Lutz}, {Netzer}, {Davies}, {Sturm}, {Pfuhl}, {Amorim}, {Baub{\"o}ck}, {Brandner}, {Cl{\'e}net}, {de Zeeuw}, {Eckart}, {Eisenhauer}, {F{\"o}rster Schreiber}, {Gao}, {Garcia}, {Genzel}, {Gillessen}, {Gratadour}, {Jim{\'e}nez-Rosales}, {Lacour}, {Millour}, {Ott}, {Paumard}, {Perraut}, {Perrin}, {Peterson}, {Petrucci}, {Prieto}, {Rouan}, {Schartmann}, {Shimizu}, {Sternberg}, {Straub}, {Straubmeier}, {Tacconi}, {Tristram}, {Vermot}, {Waisberg}, {Widmann}, \& {Woillez}}]{gravity20}
{GRAVITY Collaboration}, {Dexter}, J., {Shangguan}, J., {et~al.} 2020{\natexlab{a}}, \aap, 635, A92

\bibitem[{{GRAVITY Collaboration} {et~al.}(2020{\natexlab{b}}){GRAVITY Collaboration}, {Dexter}, {Shangguan}, {H{\"o}nig}, {Kishimoto}, {Lutz}, {Netzer}, {Davies}, {Sturm}, {Pfuhl}, {Amorim}, {Baub{\"o}ck}, {Brandner}, {Cl{\'e}net}, {de Zeeuw}, {Eckart}, {Eisenhauer}, {F{\"o}rster Schreiber}, {Gao}, {Garcia}, {Genzel}, {Gillessen}, {Gratadour}, {Jim{\'e}nez-Rosales}, {Lacour}, {Millour}, {Ott}, {Paumard}, {Perraut}, {Perrin}, {Peterson}, {Petrucci}, {Prieto}, {Rouan}, {Schartmann}, {Shimizu}, {Sternberg}, {Straub}, {Straubmeier}, {Tacconi}, {Tristram}, {Vermot}, {Waisberg}, {Widmann}, \& {Woillez}}]{dexter_gravity20}
{GRAVITY Collaboration}, {Dexter}, J., {Shangguan}, J., {et~al.} 2020{\natexlab{b}}, \aap, 635, A92

\bibitem[{{Gravity Collaboration} {et~al.}(2018){Gravity Collaboration}, {Sturm}, {Dexter}, {Pfuhl}, {Stock}, {Davies}, {Lutz}, {Cl{\'e}net}, {Eckart}, {Eisenhauer}, {Genzel}, {Gratadour}, {H{\"o}nig}, {Kishimoto}, {Lacour}, {Millour}, {Netzer}, {Perrin}, {Peterson}, {Petrucci}, {Rouan}, {Waisberg}, {Woillez}, {Amorim}, {Brandner}, {F{\"o}rster Schreiber}, {Garcia}, {Gillessen}, {Ott}, {Paumard}, {Perraut}, {Scheithauer}, {Straubmeier}, {Tacconi}, \& {Widmann}}]{Gravity18}
{Gravity Collaboration}, {Sturm}, E., {Dexter}, J., {et~al.} 2018, \nat, 563, 657

\bibitem[{{H{\"o}nig} {et~al.}(2013){H{\"o}nig}, {Kishimoto}, {Tristram}, {Prieto}, {Gandhi}, {Asmus}, {Antonucci}, {Burtscher}, {Duschl}, \& {Weigelt}}]{Honig+13}
{H{\"o}nig}, S.~F., {Kishimoto}, M., {Tristram}, K.~R.~W., {et~al.} 2013, \apj, 771, 87

\bibitem[{{Jana} {et~al.}(2025){Jana}, {Ricci}, {Temple}, {Chang}, {Shablovinskaya}, {Trakhtenbrot}, {Diaz}, {Ilic}, {Nandi}, \& {Koss}}]{Jana+25}
{Jana}, A., {Ricci}, C., {Temple}, M.~J., {et~al.} 2025, \aap, 693, A35

\bibitem[{{Jiang} {et~al.}(2022){Jiang}, {Gallo}, {Grupe}, \& {Parker}}]{Jiang22}
{Jiang}, J., {Gallo}, L.~C., {Grupe}, D., \& {Parker}, M.~L. 2022, \mnras, 516, 4256

\bibitem[{Jing {et~al.}(2025)Jing, Chen, Deng, Zhu, Zou, \& Wu}]{desi_dual}
Jing, L., Chen, Q., Deng, Z., {et~al.} 2025, A Quasar Pair Catalog Compiled from DESI DR1

\bibitem[{{Junkkarinen} {et~al.}(2001){Junkkarinen}, {Shields}, {Beaver}, {Burbidge}, {Cohen}, {Hamann}, \& {Lyons}}]{Junkkarinen01}
{Junkkarinen}, V., {Shields}, G.~A., {Beaver}, E.~A., {et~al.} 2001, \apjl, 549, L155

\bibitem[{{Kaspi} {et~al.}(2001){Kaspi}, {Brandt}, {Netzer}, {George}, {Chartas}, {Behar}, {Sambruna}, {Garmire}, \& {Nousek}}]{Kaspi01}
{Kaspi}, S., {Brandt}, W.~N., {Netzer}, H., {et~al.} 2001, \apj, 554, 216

\bibitem[{{Kelly} \& {Merloni}(2012)}]{BrandonMerloni11}
{Kelly}, B.~C. \& {Merloni}, A. 2012, Advances in Astronomy, 2012, 970858

\bibitem[{{Kollatschny} {et~al.}(2000){Kollatschny}, {Bischoff}, \& {Dietrich}}]{Kollatschny00}
{Kollatschny}, W., {Bischoff}, K., \& {Dietrich}, M. 2000, \aap, 361, 901

\bibitem[{{Kormendy} \& {Ho}(2013)}]{Kormendy_Ho_13}
{Kormendy}, J. \& {Ho}, L.~C. 2013, \araa, 51, 511

\bibitem[{{Koss} {et~al.}(2012){Koss}, {Mushotzky}, {Treister}, {Veilleux}, {Vasudevan}, \& {Trippe}}]{koss12}
{Koss}, M., {Mushotzky}, R., {Treister}, E., {et~al.} 2012, \apjl, 746, L22

\bibitem[{{Koss} {et~al.}(2018){Koss}, {Blecha}, {Bernhard}, {Hung}, {Lu}, {Trakhtenbrot}, {Treister}, {Weigel}, {Sartori}, {Mushotzky}, {Schawinski}, {Ricci}, {Veilleux}, \& {Sanders}}]{Koss2018}
{Koss}, M.~J., {Blecha}, L., {Bernhard}, P., {et~al.} 2018, \nat, 563, 214

\bibitem[{{Kova{\v{c}}evi{\'c}} {et~al.}(2020){Kova{\v{c}}evi{\'c}}, {Songsheng}, {Wang}, \& {Popovi{\'c}}}]{Kovacevic20}
{Kova{\v{c}}evi{\'c}}, A.~B., {Songsheng}, Y.-Y., {Wang}, J.-M., \& {Popovi{\'c}}, L.~{\v{C}}. 2020, \aap, 644, A88

\bibitem[{{Krause} {et~al.}(2025){Krause}, {Bourne}, {Britzen}, {Foord}, {Greene}, {Habouzit}, {Horton}, {Mayer}, {Middleton}, {Nealon}, {Sisk-Reyn{\'e}s}, {Reynolds}, \& {Sijacki}}]{Krause25}
{Krause}, M. G.~H., {Bourne}, M.~A., {Britzen}, S., {et~al.} 2025, \pasa, 42, e162

\bibitem[{{Lal} {et~al.}(2004){Lal}, {Shastri}, \& {Gabuzda}}]{Lal04}
{Lal}, D.~V., {Shastri}, P., \& {Gabuzda}, D.~C. 2004, \aap, 425, 99

\bibitem[{{Lapeyrere} {et~al.}(2014){Lapeyrere}, {Kervella}, {Lacour}, {Azouaoui}, {Garcia-Dabo}, {Perrin}, {Eisenhauer}, {Perraut}, {Straubmeier}, {Amorim}, \& {Brandner}}]{Lapeyrere14}
{Lapeyrere}, V., {Kervella}, P., {Lacour}, S., {et~al.} 2014, in Society of Photo-Optical Instrumentation Engineers (SPIE) Conference Series, Vol. 9146, Optical and Infrared Interferometry IV, ed. J.~K. {Rajagopal}, M.~J. {Creech-Eakman}, \& F.~{Malbet}, 91462D

\bibitem[{{Leethochawalit} {et~al.}(2022){Leethochawalit}, {Trenti}, {Morishita}, {Roberts-Borsani}, \& {Treu}}]{Leethochawalit22}
{Leethochawalit}, N., {Trenti}, M., {Morishita}, T., {Roberts-Borsani}, G., \& {Treu}, T. 2022, \mnras, 509, 5836

\bibitem[{{Lemon} {et~al.}(2024){Lemon}, {Courbin}, {More}, {Schechter}, {Ca{\~n}ameras}, {Delchambre}, {Leung}, {Shu}, {Spiniello}, {Hezaveh}, {Kl{\"u}ter}, \& {McMahon}}]{Lemon23}
{Lemon}, C., {Courbin}, F., {More}, A., {et~al.} 2024, \ssr, 220, 23

\bibitem[{{Lemon} {et~al.}(2017){Lemon}, {Auger}, {McMahon}, \& {Koposov}}]{Lemon17}
{Lemon}, C.~A., {Auger}, M.~W., {McMahon}, R.~G., \& {Koposov}, S.~E. 2017, \mnras, 472, 5023

\bibitem[{{Li} {et~al.}(2020{\natexlab{a}}){Li}, {Bogdanovi{\'c}}, \& {Ballantyne}}]{2020ApJ...896..113L}
{Li}, K., {Bogdanovi{\'c}}, T., \& {Ballantyne}, D.~R. 2020{\natexlab{a}}, \apj, 896, 113

\bibitem[{{Li} {et~al.}(2020{\natexlab{b}}){Li}, {Bogdanovi{\'c}}, \& {Ballantyne}}]{2020ApJ...905..123L}
{Li}, K., {Bogdanovi{\'c}}, T., \& {Ballantyne}, D.~R. 2020{\natexlab{b}}, \apj, 905, 123

\bibitem[{{Li} {et~al.}(2022){Li}, {Bogdanovi{\'c}}, {Ballantyne}, \& {Bonetti}}]{Li22}
{Li}, K., {Bogdanovi{\'c}}, T., {Ballantyne}, D.~R., \& {Bonetti}, M. 2022, \apj, 933, 104

\bibitem[{{Li} {et~al.}(2019){Li}, {Wang}, {Zhang}, {Wang}, {Huang}, {Lu}, {Hu}, {Du}, {Bon}, {Ho}, {Bai}, {Bian}, {Yuan}, {Winkler}, {Denissyuk}, {Valiullin}, {Bon}, \& {Popovi{\'c}}}]{Li17}
{Li}, Y.-R., {Wang}, J.-M., {Zhang}, Z.-X., {et~al.} 2019, \apjs, 241, 33

\bibitem[{{Liu} {et~al.}(2010){Liu}, {Greene}, {Shen}, \& {Strauss}}]{Liu10_binary_doublepeak}
{Liu}, X., {Greene}, J.~E., {Shen}, Y., \& {Strauss}, M.~A. 2010, \apjl, 715, L30

\bibitem[{{Lobban} {et~al.}(2020){Lobban}, {Zola}, {Pajdosz-{\'S}mierciak}, {Braito}, {Nardini}, {Bhatta}, {Markowitz}, {Bachev}, {Carosati}, {Caton}, {Damljanovic}, {D{\k{e}}bski}, {Haislip}, {Hu}, {Kouprianov}, {Krzesi{\'n}ski}, {Porquet}, {Pozo Nu{\~n}ez}, {Reeves}, \& {Reichart}}]{Lobban20}
{Lobban}, A.~P., {Zola}, S., {Pajdosz-{\'S}mierciak}, U., {et~al.} 2020, \mnras, 494, 1165

\bibitem[{{L{\'o}pez-Corredoira} \& {Guti{\'e}rrez}(2002)}]{LopezCorredoira02}
{L{\'o}pez-Corredoira}, M. \& {Guti{\'e}rrez}, C.~M. 2002, \aap, 390, L15

\bibitem[{{L{\'o}pez-Gonzaga} \& {Jaffe}(2016)}]{LopezGonzaga16}
{L{\'o}pez-Gonzaga}, N. \& {Jaffe}, W. 2016, \aap, 591, A128

\bibitem[{{Maiolino} {et~al.}(2024){Maiolino}, {Scholtz}, {Curtis-Lake}, {Carniani}, {Baker}, {de Graaff}, {Tacchella}, {{\"U}bler}, {D'Eugenio}, {Witstok}, {Curti}, {Arribas}, {Bunker}, {Charlot}, {Chevallard}, {Eisenstein}, {Egami}, {Ji}, {Jones}, {Lyu}, {Rawle}, {Robertson}, {Rujopakarn}, {Perna}, {Sun}, {Venturi}, {Williams}, \& {Willott}}]{Maiolino2024dual}
{Maiolino}, R., {Scholtz}, J., {Curtis-Lake}, E., {et~al.} 2024, \aap, 691, A145

\bibitem[{{Mannucci} {et~al.}(2022){Mannucci}, {Pancino}, {Belfiore}, {Cicone}, {Ciurlo}, {Cresci}, {Lusso}, {Marasco}, {Marconi}, {Nardini}, {Pinna}, {Severgnini}, {Saracco}, {Tozzi}, \& {Yeh}}]{Mannucci22}
{Mannucci}, F., {Pancino}, E., {Belfiore}, F., {et~al.} 2022, Nature Astronomy, 6, 1185

\bibitem[{{Mannucci} {et~al.}(2023){Mannucci}, {Scialpi}, {Ciurlo}, {Yeh}, {Marconcini}, {Tozzi}, {Cresci}, {Marconi}, {Amiri}, {Belfiore}, {Carniani}, {Cicone}, {Nardini}, {Pancino}, {Rubinur}, {Severgnini}, {Ulivi}, {Venturi}, {Vignali}, {Volonteri}, {Pinna}, {Rossi}, {Puglisi}, {Agapito}, {Plantet}, {Ghose}, {Carbonaro}, {Xompero}, {Grani}, {Esposito}, {Power}, {Guerra Ramon}, {Lefebvre}, {Cavallaro}, {Davies}, {Riccardi}, {Macintosh}, {Taylor}, {Dolci}, {Baruffolo}, {Feuchtgruber}, {Kravchenko}, {Rau}, {Sturm}, {Wiezorrek}, {Dallilar}, \& {Kenworthy}}]{Mannucci23}
{Mannucci}, F., {Scialpi}, M., {Ciurlo}, A., {et~al.} 2023, \aap, 680, A53

\bibitem[{{Matsuoka} {et~al.}(2024){Matsuoka}, {Izumi}, {Onoue}, {Strauss}, {Iwasawa}, {Kashikawa}, {Akiyama}, {Aoki}, {Arita}, {Imanishi}, {Ishimoto}, {Kawaguchi}, {Kohno}, {Lee}, {Nagao}, {Silverman}, \& {Toba}}]{Matsuoka2024}
{Matsuoka}, Y., {Izumi}, T., {Onoue}, M., {et~al.} 2024, \apjl, 965, L4

\bibitem[{{Mehdipour} {et~al.}(2017){Mehdipour}, {Kaastra}, {Kriss}, {Arav}, {Behar}, {Bianchi}, {Branduardi-Raymont}, {Cappi}, {Costantini}, {Ebrero}, {Di Gesu}, {Kaspi}, {Mao}, {De Marco}, {Matt}, {Paltani}, {Peretz}, {Peterson}, {Petrucci}, {Pinto}, {Ponti}, {Ursini}, {de Vries}, \& {Walton}}]{Mehdipour17}
{Mehdipour}, M., {Kaastra}, J.~S., {Kriss}, G.~A., {et~al.} 2017, \aap, 607, A28

\bibitem[{{M{\'e}rand}(2022)}]{PMOIRED22}
{M{\'e}rand}, A. 2022, in Society of Photo-Optical Instrumentation Engineers (SPIE) Conference Series, Vol. 12183, Optical and Infrared Interferometry and Imaging VIII, ed. A.~{M{\'e}rand}, S.~{Sallum}, \& J.~{Sanchez-Bermudez}, 121831N

\bibitem[{{Mezcua} {et~al.}(2026){Mezcua}, {Laloux}, {Scialpi}, {Siudek}, {Er{\'o}stegui}, {Ricci}, {Matamoro Zatarain}, {Visser}, {Rottgering}, {Gutierrez}, {Feltre}, {Bisigello}, \& {Saulder}}]{mezcua26}
{Mezcua}, M., {Laloux}, B., {Scialpi}, M., {et~al.} 2026, arXiv e-prints, arXiv:2604.13170

\bibitem[{{Monnier}(2003)}]{Monnier03}
{Monnier}, J.~D. 2003, in EAS Publications Series, Vol.~6, EAS Publications Series, ed. G.~{Perrin} \& F.~{Malbet} (EDP), 213

\bibitem[{{M{\"u}ller-S{\'a}nchez} {et~al.}(2011){M{\"u}ller-S{\'a}nchez}, {Prieto}, {Hicks}, {Vives-Arias}, {Davies}, {Malkan}, {Tacconi}, \& {Genzel}}]{MullerSanchez11}
{M{\"u}ller-S{\'a}nchez}, F., {Prieto}, M.~A., {Hicks}, E.~K.~S., {et~al.} 2011, \apj, 739, 69

\bibitem[{{Perna} {et~al.}(2025{\natexlab{a}}){Perna}, {Arribas}, {Lamperti}, {Circosta}, {Bertola}, {P{\'e}rez-Gonz{\'a}lez}, {D'Eugenio}, {{\"U}bler}, {Cresci}, {Volonteri}, {Mannucci}, {Maiolino}, {Rodr{\'\i}guez Del Pino}, {B{\"o}ker}, {Bunker}, {Charlot}, {Willott}, {Carniani}, {Curti}, {Jones}, {Kumari}, {Marshall}, {Venturi}, {Saxena}, {Scholtz}, \& {Witstok}}]{Perna2025dual}
{Perna}, M., {Arribas}, S., {Lamperti}, I., {et~al.} 2025{\natexlab{a}}, \aap, 696, A59

\bibitem[{{Perna} {et~al.}(2025{\natexlab{b}}){Perna}, {Arribas}, {Lamperti}, {Circosta}, {Bertola}, {P{\'e}rez-Gonz{\'a}lez}, {D'Eugenio}, {{\"U}bler}, {Cresci}, {Volonteri}, {Mannucci}, {Maiolino}, {Rodr{\'\i}guez Del Pino}, {B{\"o}ker}, {Bunker}, {Charlot}, {Willott}, {Carniani}, {Curti}, {Jones}, {Kumari}, {Marshall}, {Venturi}, {Saxena}, {Scholtz}, \& {Witstok}}]{Perna2025}
{Perna}, M., {Arribas}, S., {Lamperti}, I., {et~al.} 2025{\natexlab{b}}, \aap, 696, A59

\bibitem[{{Planck Collaboration} {et~al.}(2020){Planck Collaboration}, {Aghanim}, {Akrami}, {Ashdown}, {Aumont}, {Baccigalupi}, {Ballardini}, {Banday}, {Barreiro}, {Bartolo}, {Basak}, {Battye}, {Benabed}, {Bernard}, {Bersanelli}, {Bielewicz}, {Bock}, {Bond}, {Borrill}, {Bouchet}, {Boulanger}, {Bucher}, {Burigana}, {Butler}, {Calabrese}, {Cardoso}, {Carron}, {Challinor}, {Chiang}, {Chluba}, {Colombo}, {Combet}, {Contreras}, {Crill}, {Cuttaia}, {de Bernardis}, {de Zotti}, {Delabrouille}, {Delouis}, {Di Valentino}, {Diego}, {Dor{\'e}}, {Douspis}, {Ducout}, {Dupac}, {Dusini}, {Efstathiou}, {Elsner}, {En{\ss}lin}, {Eriksen}, {Fantaye}, {Farhang}, {Fergusson}, {Fernandez-Cobos}, {Finelli}, {Forastieri}, {Frailis}, {Fraisse}, {Franceschi}, {Frolov}, {Galeotta}, {Galli}, {Ganga}, {G{\'e}nova-Santos}, {Gerbino}, {Ghosh}, {Gonz{\'a}lez-Nuevo}, {G{\'o}rski}, {Gratton}, {Gruppuso}, {Gudmundsson}, {Hamann}, {Handley}, {Hansen}, {Herranz}, {Hildebrandt}, {Hivon}, {Huang}, {Jaffe}, {Jones}, {Karakci}, {Keih{\"a}nen},
  {Keskitalo}, {Kiiveri}, {Kim}, {Kisner}, {Knox}, {Krachmalnicoff}, {Kunz}, {Kurki-Suonio}, {Lagache}, {Lamarre}, {Lasenby}, {Lattanzi}, {Lawrence}, {Le Jeune}, {Lemos}, {Lesgourgues}, {Levrier}, {Lewis}, {Liguori}, {Lilje}, {Lilley}, {Lindholm}, {L{\'o}pez-Caniego}, {Lubin}, {Ma}, {Mac{\'\i}as-P{\'e}rez}, {Maggio}, {Maino}, {Mandolesi}, {Mangilli}, {Marcos-Caballero}, {Maris}, {Martin}, {Martinelli}, {Mart{\'\i}nez-Gonz{\'a}lez}, {Matarrese}, {Mauri}, {McEwen}, {Meinhold}, {Melchiorri}, {Mennella}, {Migliaccio}, {Millea}, {Mitra}, {Miville-Desch{\^e}nes}, {Molinari}, {Montier}, {Morgante}, {Moss}, {Natoli}, {N{\o}rgaard-Nielsen}, {Pagano}, {Paoletti}, {Partridge}, {Patanchon}, {Peiris}, {Perrotta}, {Pettorino}, {Piacentini}, {Polastri}, {Polenta}, {Puget}, {Rachen}, {Reinecke}, {Remazeilles}, {Renzi}, {Rocha}, {Rosset}, {Roudier}, {Rubi{\~n}o-Mart{\'\i}n}, {Ruiz-Granados}, {Salvati}, {Sandri}, {Savelainen}, {Scott}, {Shellard}, {Sirignano}, {Sirri}, {Spencer}, {Sunyaev}, {Suur-Uski}, {Tauber}, {Tavagnacco},
  {Tenti}, {Toffolatti}, {Tomasi}, {Trombetti}, {Valenziano}, {Valiviita}, {Van Tent}, {Vibert}, {Vielva}, {Villa}, {Vittorio}, {Wandelt}, {Wehus}, {White}, {White}, {Zacchei}, \& {Zonca}}]{Planck2020}
{Planck Collaboration}, {Aghanim}, N., {Akrami}, Y., {et~al.} 2020, \aap, 641, A6

\bibitem[{{Power} {et~al.}(2003){Power}, {Navarro}, {Jenkins}, {Frenk}, {White}, {Springel}, {Stadel}, \& {Quinn}}]{poweretal03}
{Power}, C., {Navarro}, J.~F., {Jenkins}, A., {et~al.} 2003, \mnras, 338, 14

\bibitem[{{Quelquejay Leclere} {et~al.}(2025){Quelquejay Leclere}, {Li}, {Volonteri}, {Babak}, {Beckmann}, {Dubois}, {Laigle}, \& {Webb}}]{Hippolyte2025}
{Quelquejay Leclere}, H., {Li}, K., {Volonteri}, M., {et~al.} 2025, arXiv e-prints, arXiv:2510.14613

\bibitem[{{Reeves} {et~al.}(2004){Reeves}, {Nandra}, {George}, {Pounds}, {Turner}, \& {Yaqoob}}]{Reeves04}
{Reeves}, J.~N., {Nandra}, K., {George}, I.~M., {et~al.} 2004, \apj, 602, 648

\bibitem[{{Rodriguez} {et~al.}(2006){Rodriguez}, {Taylor}, {Zavala}, {Peck}, {Pollack}, \& {Romani}}]{Rodriguez06}
{Rodriguez}, C., {Taylor}, G.~B., {Zavala}, R.~T., {et~al.} 2006, \apj, 646, 49

\bibitem[{{Rosas-Guevara} {et~al.}(2019){Rosas-Guevara}, {Bower}, {McAlpine}, {Bonoli}, \& {Tissera}}]{Rosas-Guevara2019}
{Rosas-Guevara}, Y.~M., {Bower}, R.~G., {McAlpine}, S., {Bonoli}, S., \& {Tissera}, P.~B. 2019, \mnras, 483, 2712

\bibitem[{{Runnoe} {et~al.}(2025){Runnoe}, {Eracleous}, {Bogdanovi{\'c}}, {Halpern}, \& {Sigur{\dj}sson}}]{Runnoe25}
{Runnoe}, J.~C., {Eracleous}, M., {Bogdanovi{\'c}}, T., {Halpern}, J.~P., \& {Sigur{\dj}sson}, S. 2025, \apj, 984, 17

\bibitem[{{Runnoe} {et~al.}(2017){Runnoe}, {Eracleous}, {Pennell}, {Mathes}, {Boroson}, {Sigur{\dh}sson}, {Bogdanovi{\'c}}, {Halpern}, {Liu}, \& {Brown}}]{Runnoe17}
{Runnoe}, J.~C., {Eracleous}, M., {Pennell}, A., {et~al.} 2017, \mnras, 468, 1683

\bibitem[{{Saeedzadeh} {et~al.}(2024){Saeedzadeh}, {Babul}, {Mukherjee}, {Tremmel}, {Quinn}, \& {Mayer}}]{Saeedzadeh23_romulus_dual}
{Saeedzadeh}, V., {Babul}, A., {Mukherjee}, S., {et~al.} 2024, \apj, 975, 265

\bibitem[{{Sayeb} {et~al.}(2021){Sayeb}, {Blecha}, {Kelley}, {Gerosa}, {Kesden}, \& {Thomas}}]{Kelley21}
{Sayeb}, M., {Blecha}, L., {Kelley}, L.~Z., {et~al.} 2021, \mnras, 501, 2531

\bibitem[{{Schulze} \& {Wisotzki}(2010)}]{Schulze10}
{Schulze}, A. \& {Wisotzki}, L. 2010, \aap, 516, A87

\bibitem[{{Scialpi} {et~al.}(2025){Scialpi}, {Mannucci}, {D'Amato}, {Marconcini}, {Cresci}, {Marconi}, {Ulivi}, {Fumagalli}, {Rosati}, {Tozzi}, {Zanchettin}, {Cataldi}, {Battistini}, {Bertola}, {Bracci}, {Carniani}, {Ceci}, {Chakraborty}, {Cicone}, {Ciurlo}, {De Rosa}, {Di Rosa}, {Feltre}, {Ginolfi}, {Lamperti}, {Lusso}, {Moreschini}, {Nardini}, {Parvatikar}, {Perna}, {Rubinur}, {Severgnini}, {Singh}, {Spingola}, {Venturi}, {Vignali}, \& {Volonteri}}]{scialpi25_arxive}
{Scialpi}, M., {Mannucci}, F., {D'Amato}, Q., {et~al.} 2025, arXiv e-prints, arXiv:2512.11960

\bibitem[{{Scialpi} {et~al.}(2024){Scialpi}, {Mannucci}, {Marconcini}, {Venturi}, {Pancino}, {Marconi}, {Cresci}, {Belfiore}, {Amiri}, {Bertola}, {Carniani}, {Cicone}, {Ciurlo}, {D'Amato}, {Ginolfi}, {Lusso}, {Marasco}, {Nardini}, {Rubinur}, {Severgnini}, {Tozzi}, {Ulivi}, {Vignali}, \& {Volonteri}}]{Scialpi24}
{Scialpi}, M., {Mannucci}, F., {Marconcini}, C., {et~al.} 2024, \aap, 690, A57

\bibitem[{{Severgnini} {et~al.}(2018){Severgnini}, {Cicone}, {Della Ceca}, {Braito}, {Caccianiga}, {Ballo}, {Campana}, {Moretti}, {La Parola}, {Vignali}, {Zaino}, {Matzeu}, \& {Landoni}}]{Severgnini18_binary_doublepeak}
{Severgnini}, P., {Cicone}, C., {Della Ceca}, R., {et~al.} 2018, \mnras, 479, 3804

\bibitem[{{Shen} {et~al.}(2020){Shen}, {Hopkins}, {Faucher-Gigu{\`e}re}, {Alexander}, {Richards}, {Ross}, \& {Hickox}}]{2020MNRAS.495.3252S}
{Shen}, X., {Hopkins}, P.~F., {Faucher-Gigu{\`e}re}, C.-A., {et~al.} 2020, \mnras, 495, 3252

\bibitem[{{Shen} {et~al.}(2023){Shen}, {Hwang}, {Oguri}, {Chen}, {Di Matteo}, {Ni}, {Bird}, {Zakamska}, {Liu}, {Chen}, \& {Kratter}}]{Shen23b}
{Shen}, Y., {Hwang}, H.-C., {Oguri}, M., {et~al.} 2023, \apj, 943, 38

\bibitem[{{Songsheng} {et~al.}(2019){Songsheng}, {Wang}, {Li}, \& {Du}}]{Songsheng19}
{Songsheng}, Y.-Y., {Wang}, J.-M., {Li}, Y.-R., \& {Du}, P. 2019, \apj, 881, 140

\bibitem[{{Tohline} \& {Osterbrock}(1976)}]{Tohline76}
{Tohline}, J.~E. \& {Osterbrock}, D.~E. 1976, \apjl, 210, L117

\bibitem[{{Trippe} {et~al.}(2010){Trippe}, {Crenshaw}, {Deo}, {Dietrich}, {Kraemer}, {Rafter}, \& {Turner}}]{Trippe10}
{Trippe}, M.~L., {Crenshaw}, D.~M., {Deo}, R.~P., {et~al.} 2010, \apj, 725, 1749

\bibitem[{{{\"U}bler} {et~al.}(2024){{\"U}bler}, {Maiolino}, {P{\'e}rez-Gonz{\'a}lez}, {D'Eugenio}, {Perna}, {Curti}, {Arribas}, {Bunker}, {Carniani}, {Charlot}, {Rodr{\'\i}guez Del Pino}, {Baker}, {B{\"o}ker}, {Cresci}, {Dunlop}, {Grogin}, {Jones}, {Kumari}, {Lamperti}, {Laporte}, {Marshall}, {Mazzolari}, {Parlanti}, {Rawle}, {Scholtz}, {Venturi}, \& {Witstok}}]{Ubler2024dual}
{{\"U}bler}, H., {Maiolino}, R., {P{\'e}rez-Gonz{\'a}lez}, P.~G., {et~al.} 2024, \mnras, 531, 355

\bibitem[{{Ulivi} {et~al.}(2025){Ulivi}, {Mannucci}, {Scialpi}, {Marconcini}, {Cresci}, {Marconi}, {Feltre}, \& {Ginolfi}}]{Ulivi2025arXiv}
{Ulivi}, L., {Mannucci}, F., {Scialpi}, M., {et~al.} 2025, arXiv e-prints, arXiv:2508.19494

\bibitem[{{Volonteri} {et~al.}(2016){Volonteri}, {Dubois}, {Pichon}, \& {Devriendt}}]{Volonteri16}
{Volonteri}, M., {Dubois}, Y., {Pichon}, C., \& {Devriendt}, J. 2016, \mnras, 460, 2979

\bibitem[{{Volonteri} {et~al.}(2003){Volonteri}, {Haardt}, \& {Madau}}]{Volonteri03}
{Volonteri}, M., {Haardt}, F., \& {Madau}, P. 2003, \apj, 582, 559

\bibitem[{{Volonteri} {et~al.}(2022){Volonteri}, {Pfister}, {Beckmann}, {Dotti}, {Dubois}, {Massonneau}, {Musoke}, \& {Tremmel}}]{Volonteri22}
{Volonteri}, M., {Pfister}, H., {Beckmann}, R., {et~al.} 2022, \mnras, 514, 640

\bibitem[{{Volonteri} {et~al.}(2020){Volonteri}, {Pfister}, {Beckmann}, {Dubois}, {Colpi}, {Conselice}, {Dotti}, {Martin}, {Jackson}, {Kraljic}, {Pichon}, {Trebitsch}, {Yi}, {Devriendt}, \& {Peirani}}]{2020MNRAS.498.2219V}
{Volonteri}, M., {Pfister}, H., {Beckmann}, R.~S., {et~al.} 2020, \mnras, 498, 2219

\bibitem[{{Wang} {et~al.}(2009){Wang}, {Chen}, {Hu}, {Mao}, {Zhang}, \& {Bian}}]{Wang09_binary_doublepeak}
{Wang}, J.-M., {Chen}, Y.-M., {Hu}, C., {et~al.} 2009, \apjl, 705, L76

\bibitem[{{White} \& {Rees}(1978)}]{White78}
{White}, S.~D.~M. \& {Rees}, M.~J. 1978, \mnras, 183, 341

\bibitem[{{Zamora} {et~al.}(2025){Zamora}, {Venturi}, {Carniani}, {Bertola}, {Parlanti}, {Perna}, {Arribas}, {B{\"o}ker}, {Bunker}, {Charlot}, {D'Eugenio}, {Maiolino}, {Del Pino}, {{\"U}bler}, {Cresci}, {Jones}, \& {Lamperti}}]{Zamora25}
{Zamora}, S., {Venturi}, G., {Carniani}, S., {et~al.} 2025, \aap, 702, A102

\end{thebibliography}


\appendix
\section{Simulated closure-phase signatures of binary AGN}
\label{app:closure_phase_simulations}

To investigate how binary AGN configurations affect the interferometric closure phase, we performed a set of simulations using two-component brightness distributions. We considered different combinations of unresolved (point-like) and resolved Gaussian components, including two point sources, one Gaussian plus one point source, and two Gaussian components. For the Gaussian components, we explored different sizes and ellipticities.

The two components were placed at projected separations of 5--30 mas, with the secondary source located at eight different position angles around the primary. We also varied the flux ratio between the two components, $F_2/F_1$, over the range $0.1$--$1$. For each configuration, the model brightness distribution was Fourier transformed and sampled at the $(u,v)$ coordinates corresponding to the observations, allowing us to compute the complex visibilities and the resulting closure phases as a function of spatial frequency.

\begin{figure*}[t]
    \centering

    \includegraphics[
        width=0.88\linewidth
    ]{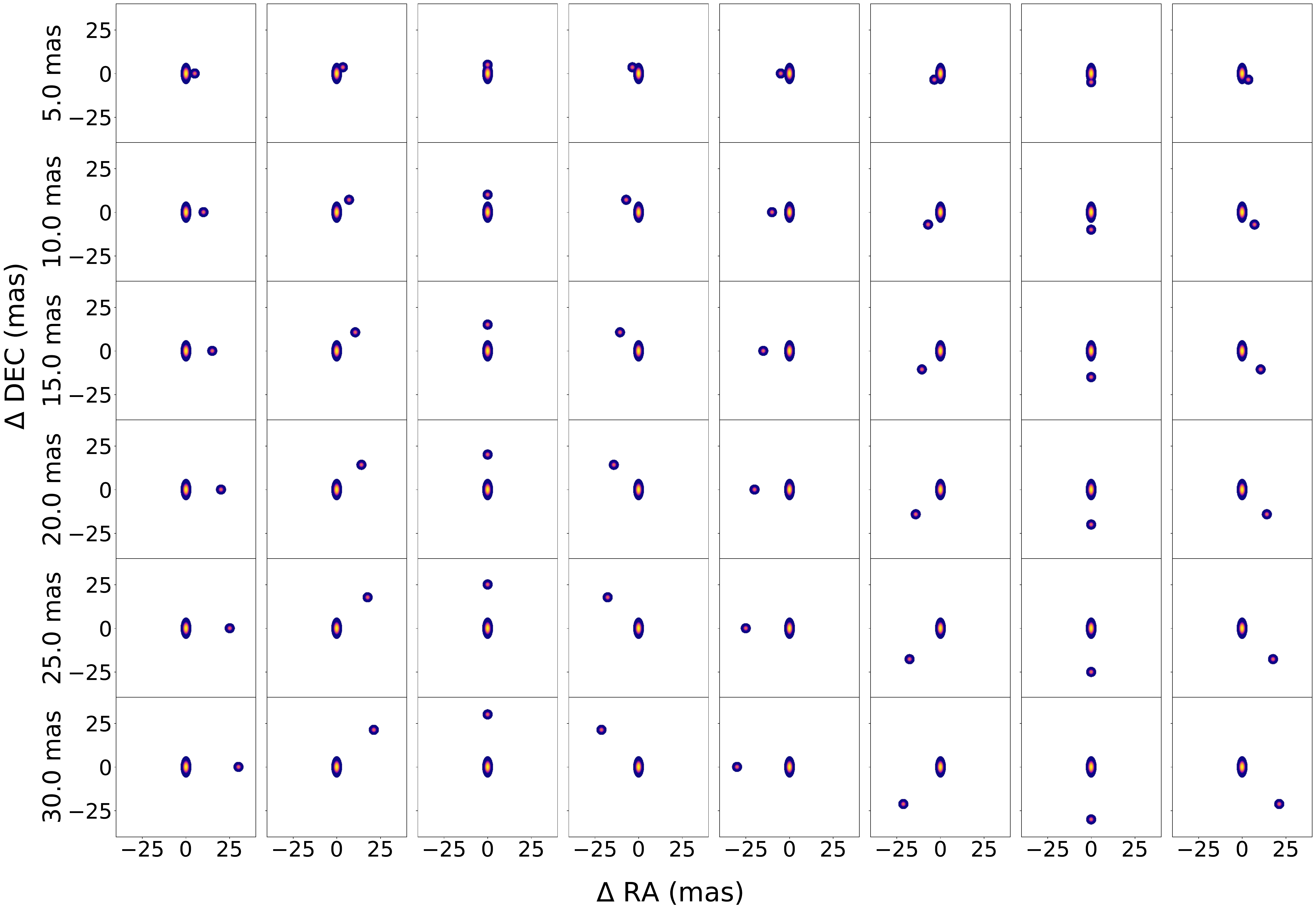}

    \vspace{0.5cm}

    \includegraphics[
        width=0.88\linewidth]{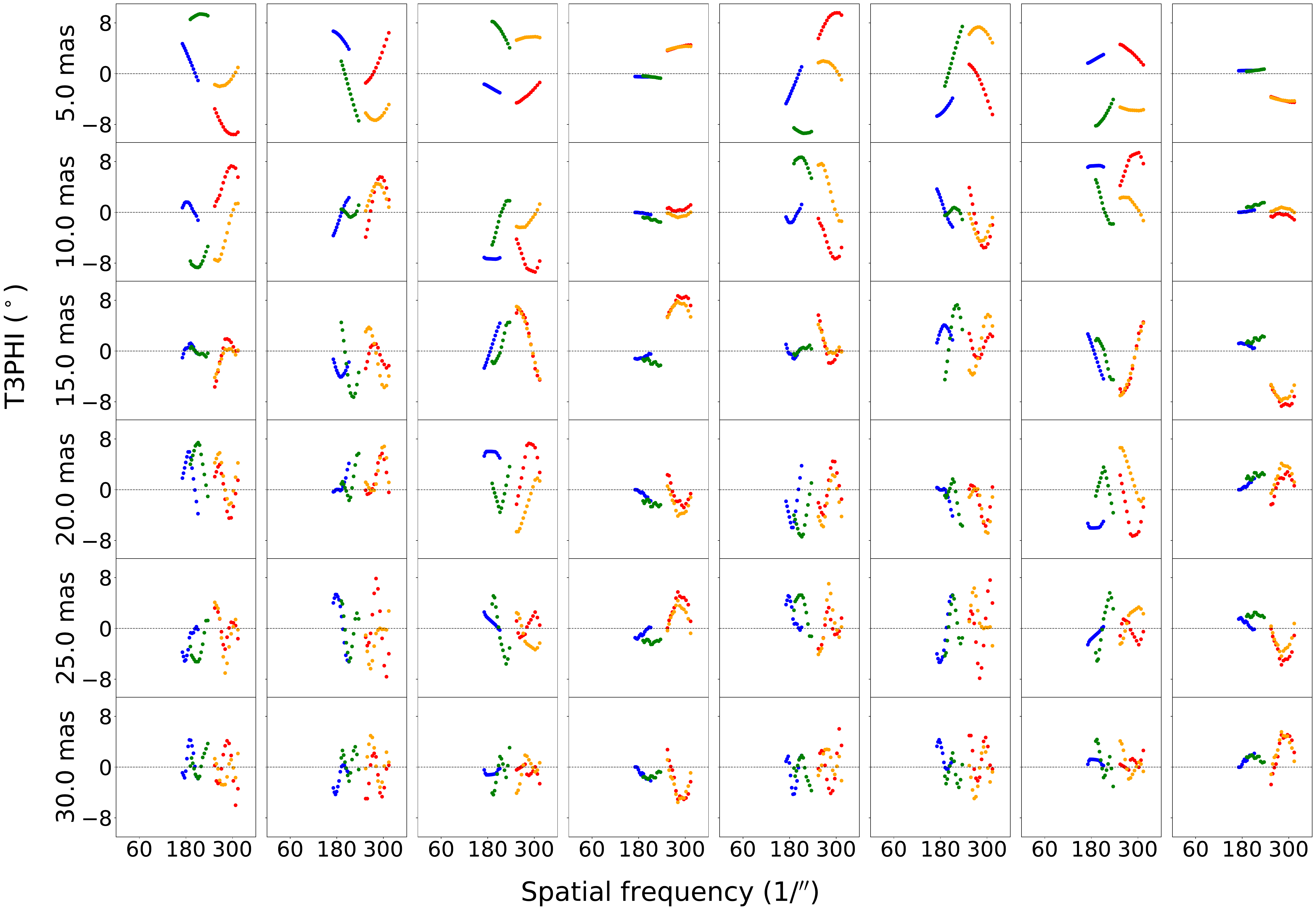}

\caption{
Simulated interferometric signatures of a two-Gaussian model with a central component characterized by $\sigma_x=0.5$~mas and $\sigma_y=1$~mas and a secondary component with $\sigma_x=\sigma_y=0.5$~mas, with $F_2/F_1=0.1$. Upper and lower panels show the brightness distributions and closure phases, respectively. Rows span projected separations of 5--35~mas, while columns show eight position angles over $360^\circ$. The four closure-phase colors correspond to the four baseline triangles.
}

    \label{fig:simul_multi_CP}
\end{figure*}

These simulations were used to investigate how the closure-phase signal depends on the main properties of a two-component system, namely the projected separation, relative position angle, flux ratio, and, for resolved components, their characteristic size and shape. In general, increasing the separation produces stronger and more rapidly varying closure-phase signatures as a function of spatial frequency, while increasing the flux ratio generally enhances the amplitude of the signal for unequal-flux configurations. For the limiting case of an equal-flux binary, however, the brightness distribution becomes centro-symmetric and the closure phase is zero. The detailed shape and amplitude of the closure phase also depend on the relative position angle and on the spatial extent and morphology of the two components.
An example of the simulated brightness distributions and corresponding closure-phase signatures is shown in Fig.~\ref{fig:simul_multi_CP}, for a two-Gaussian model with $F_2/F_1=0.1$, using the VLTI/GRAVITY telescope configuration and its four corresponding baseline triangles, shown in different colors.

\newpage
\section{Binary fit of NGC 3783}
\label{app:binaryfit}

To model the interferometric data of NGC~3783, we fit the candidate identified through the grid search following the PSF and injection-recovery validation procedures (see Section~\ref{sec:binary_candidates} for details).
The system is described by a composite brightness distribution consisting of a central Gaussian disk with an optimized FWHM, a resolved background source, and a secondary point source.

\begin{figure*}[b]
    \centering
    \includegraphics[width=\linewidth]{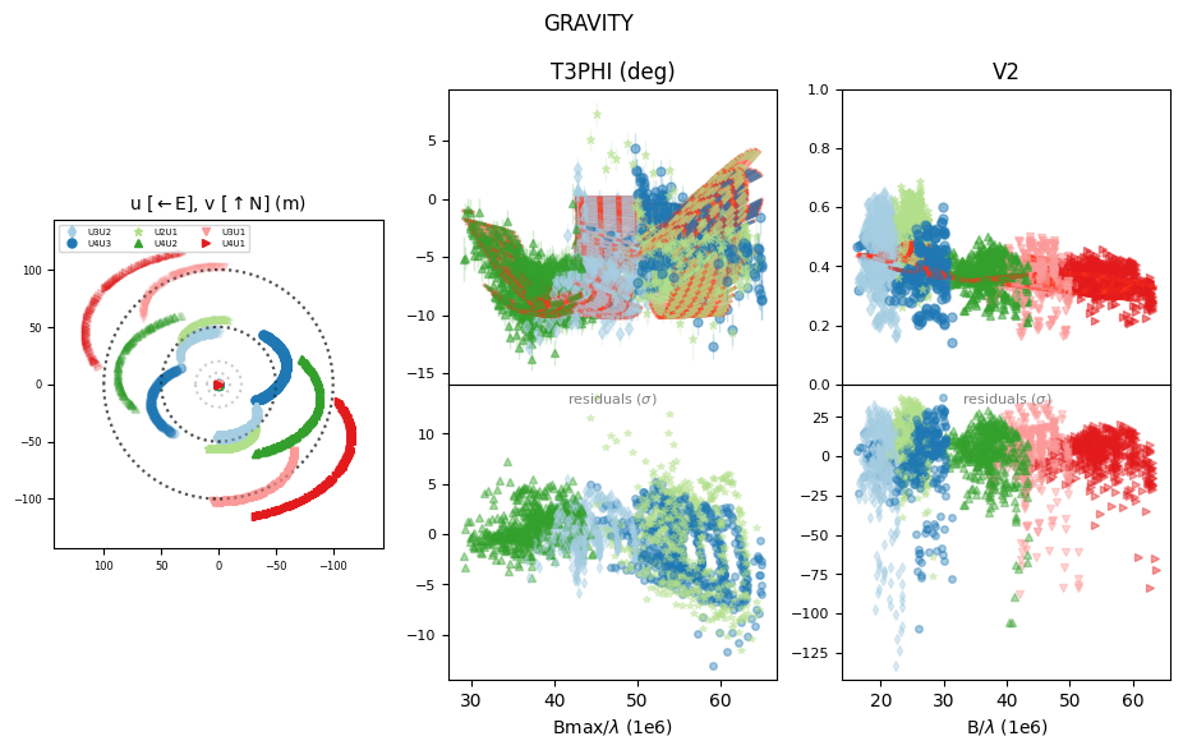}
    \caption{Model fit and residuals for NGC~3783 using a central Gaussian disk, a resolved background source, and a secondary point source.
    Left panel: Coverage of the $(u,v)$ plane in meters ($u$ East, $v$ North). Different markers and colors correspond to individual telescope baselines.
    Middle panel: Closure phase in degree(T3PHI, top) and corresponding fit residuals in units of standard deviation (bottom).
    Right panel: Squared visibility ($V2$, top) and corresponding fit residuals (bottom). The residuals are normalized by the formal statistical uncertainties provided by the GRAVITY data reduction.}
    \label{fig:fitcpv_NGC3783}
\end{figure*}

The resulting model fit and its corresponding residuals are presented in Fig.~\ref{fig:fitcpv_NGC3783}, complementing the posterior distributions shown in Fig.~\ref{fig:gaussngc3783}. The left panel displays the spatial frequency coverage in the $(u,v)$ plane. The middle and right panels show the data alongside the best-fit model for the closure phases and squared visibilities ($V^2$), respectively, together with their normalized residuals.
The residuals are expressed in units of the formal statistical uncertainties provided by the GRAVITY data reduction. For these high-S/N data, the formal uncertainties can be very small, particularly for $V^2$, such that relatively small absolute differences between the data and the model can result in large normalized residuals. These $\sigma$-residuals should therefore be interpreted relative to the formal statistical uncertainties and may also reflect systematic calibration uncertainties that are not fully captured by the pipeline errors. The model provides a good representation of the observed interferometric quantities, with the remaining discrepancies being small in absolute terms.

\end{document}